\documentclass[reprint,
superscriptaddress,
 amsmath,amssymb,
 aps,
]{revtex4-2}

\usepackage{graphicx}% Include figure files
\usepackage{dcolumn}% Align table columns on decimal point
\usepackage{bm}% bold math
\usepackage{braket}
\usepackage{hyperref}% add hypertext capabilities
\hypersetup{colorlinks = true, allcolors = blue}
\begin{document}

\preprint{APS/123-QED}

\title{Quantized topological energy pumping and Weyl points in Floquet synthetic dimensions with a driven-dissipative photonic molecule}

\author{Sashank Kaushik Sridhar}
 %\email{sashankk@umd.edu}
 \affiliation{Department of Mechanical Engineering, University of Maryland, College Park, Maryland 20742, USA}%Lines break automatically or can be forced with \\
\author{Sayan Ghosh}%
\affiliation{%
 Department of Physical Sciences, Indian Institute of Science Education and Research Kolkata,\\Mohanpur, West Bengal 741246, India
}%

% \collaboration{MUSO Collaboration}%\noaffiliation
\author{Avik Dutt}
 \affiliation{Department of Mechanical Engineering, University of Maryland, College Park, Maryland 20742, USA}
\affiliation{
 Institute for Physical Science and Technology, University of Maryland, College Park, Maryland 20742, USA
}%

\date{\today}% It is always \today, today,

\begin{abstract}
% Spins subjected to multifrequency drives show behavior analogous to $n$-D lattices under a two band approximation, where $n$ is the number of incommensurate frequencies in the drive. The topology of these bands have been studied, with proposals to realize the half-BHZ model with two Floquet synthetic dimensions, leading to analogous topological phenomena such as topological energy pumping. In practice, considerations of external drive and dissipation are crucial, since classical and quantum spins are both subject to various decay processes. Here, we introduce topological energy pumping in a quasiperiodically-modulated photonic molecule with two Floquet synthetic dimensions.
Topological effects manifest in a wide range of physical systems, such as solid crystals, acoustic waves, photonic materials and cold atoms. These effects are characterized by `topological invariants' which are typically integer-valued, and lead to robust quantized channels of transport in space, time, and other degrees of freedom. The temporal channel, in particular, allows one to achieve higher-dimensional topological effects, by driving the system with multiple incommensurate frequencies. However, dissipation is generally detrimental to such topological effects, particularly when the systems consist of quantum spins or qubits. 
Here we introduce a photonic molecule subjected to multiple RF/optical drives and dissipation as a promising candidate system to observe quantized transport along Floquet synthetic dimensions. Topological energy pumping in the incommensurately modulated photonic molecule is enhanced by the driven-dissipative nature of our platform.
Furthermore, we provide a path to realizing Weyl points and measuring the Berry curvature emanating from these reciprocal-space ($k$-space) magnetic monopoles, illustrating the capabilities for higher-dimensional topological Hamiltonian simulation in this platform. Our approach enables direct $k$-space engineering of a wide variety of Hamiltonians using modulation bandwidths that are well below the free-spectral range (FSR) of integrated photonic cavities.

\end{abstract}


\maketitle

Quantized transport is a hallmark of topological insulators (TIs). Initially explored for electronic transport in condensed matter systems (i.e. quantized Hall conductance \cite{klitzing_new_1980, thouless_quantized_1982}, Hamiltonians supporting nontrivial topology have now been experimentally simulated in a wide variety of systems such as ultracold atoms \cite{miyake_realizing_2013, aidelsburger_realization_2013}, photonics \cite{price_roadmap_2022}, acoustics \cite{khanikaev_topologically_2015, ma_topological_2019} and topolectrical circuits \cite{imhof_topolectrical-circuit_2018}. In such simulators, the Hamiltonians are typically created by controlling the coupling between a lattice of real-space sites encoded, for example, in large arrays of atoms, photonic resonators or photonic waveguides \cite{hafezi_imaging_2013, rechtsman_photonic_2013, aidelsburger_measuring_2015}. Experiments have reported robust unidirectional edge states \cite{wang_observation_2009, hafezi_imaging_2013, atala_observation_2014} in these real-space emulators, but the theoretical and experimental evidence for \textit{quantized} topological transport in photonic TIs remains scant -- especially when compared to the near-perfect quantized conductivity \cite{thouless_quantized_1982} in electronic quantum Hall systems to one part in $10^9$, which was later used to define the resistance standard \cite{KvK}.
A prime reason for this is the difficulty in defining transport properties and the analog of conductivity for neutral particles such as atoms or photons as they do not naturally respond to electromagnetic fields. Other impediments to ideal transport quantization in real-space simulators include the inescapable effects of dissipation and external driving \cite{ozawa_anomalous_2014}. 

In recent years, the concept of synthetic dimensions has emerged by repurposing internal degrees of freedom of atoms and photons as extra dimensions, thus realizing high-dimensional topological phenomena on compact, low-dimensional physical structures \cite{yuan_synthetic_2021, ozawa_topological_2019}. Synthetic dimensions have enabled lattice Hamiltonians with straightforward reconfigurability and tunability, long-range coupling, and artificial magnetic gauge fields for neutral particles, through precise control of coupling between modes labeled by degrees of freedom such as frequency, temporal modes, orbital angular momentum, spin and transverse spatial modes \cite{dutt_experimental_2019, leefmans_topological_2022, bartlett_deterministic_2021, lustig_photonic_2019, dutt_single_2020, boada_quantum_2012, stuhl_visualizing_2015, mancini_observation_2015, sundar_synthetic_2018, luo_quantum_2015, kanungo_realizing_2022, yang_realization_2023, hu_realization_2020}. Photonic synthetic \textit{frequency} dimensions, in particular, have successfully demonstrated both Hermitian and non-Hermitian topology, electromagnetic gauge fields, unidirectional edge states, Bloch oscillations, and bulk as well as boundary phenomena \cite{dutt_experimental_2019, dutt_single_2020, yuan_photonic_2016, yuan_bloch_2016, yuan_synthetic_2021, wang_generating_2021, wang_topological_2021, wang_multidimensional_2018, dutt_creating_2022, lu_universal_2020}. However, the aforementioned limitations of \textit{quantized} topological transport in real-space photonic emulators -- that of neutral particle transport in electromagnetic fields, the presence of dissipation and drive -- also apply to synthetic-space systems. 

\begin{figure}
    \centering
    \includegraphics[width = 8cm]{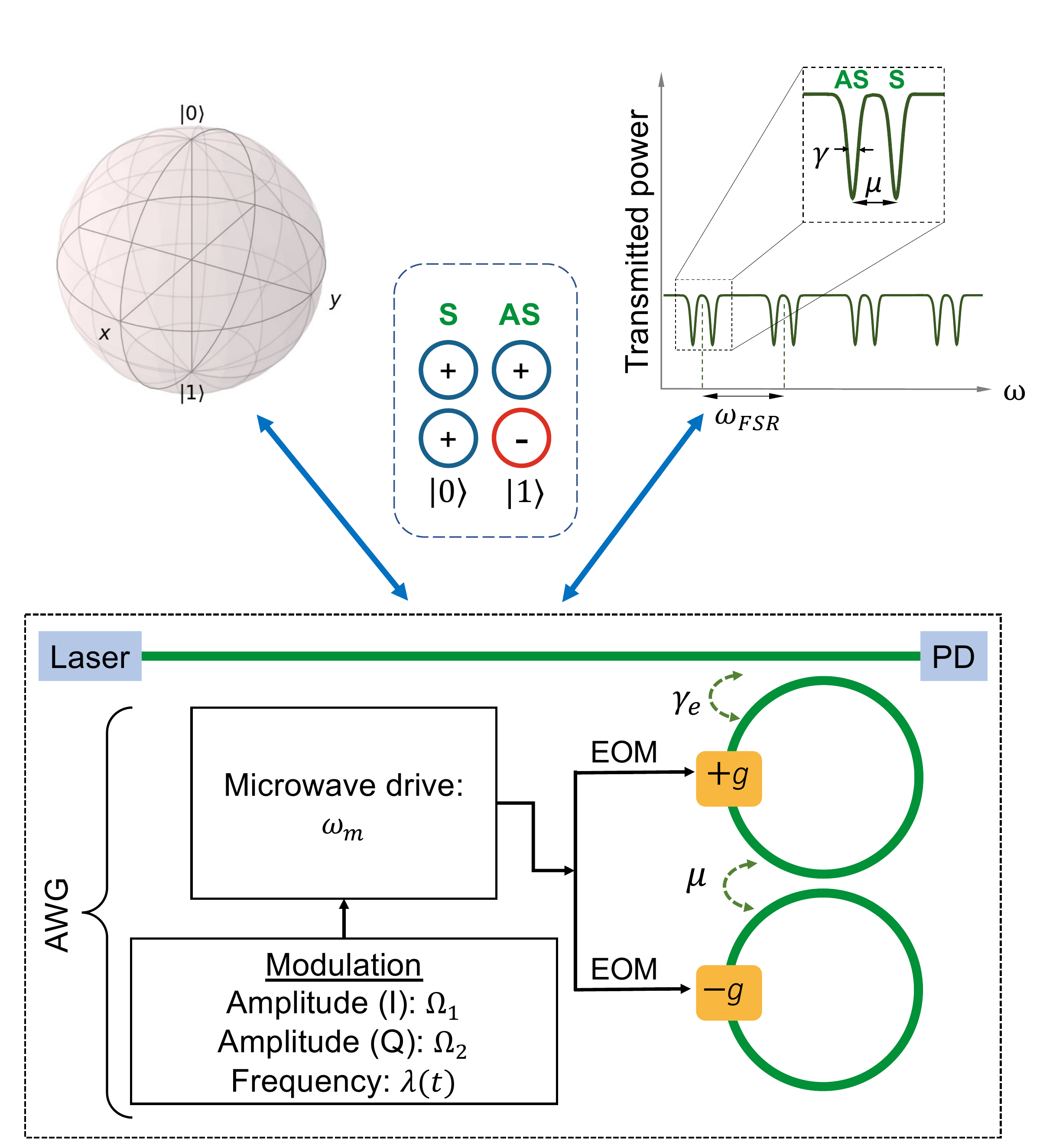}
    \caption{Schematic of the proposed system. The non-degenerate eigenmodes separated by the coupling rate $\mu$ of the photonic molecule are the symmetric (S) and anti-symmetric (AS) supermodes, i.e., single-ring azimuthal modes that are in-phase (S) or $\pi$-phase separated (AS). Coherent evolution in the subspace of a single pair of these resonance modes can be equated to the motion on a Bloch sphere, which is controlled by driving the electro-optic modulator (EOM) with an RF signal near resonance. The transmitted power at the photodetector (PD) is shown on the top right for the photonic molecule without any RF drive on the EOMs. Due to the finite resonance linewidths, the symmetric decay in mode amplitudes is given by $\gamma$. We can generate an RF signal with amplitude I-Q and frequency modulation, as defined in Eq.~\ref{eq:mod}, using an arbitrary waveform generator (AWG), allowing us to implement the half-BHZ Hamiltonian on a Floquet lattice [Eq.~\eqref{eq:BHZ}].}
    \label{fig:Schematic}
\end{figure}

Here we show how the effects of multiple drives and dissipation can support quantized topological transport of photons, by using the concept of Floquet synthetic dimensions in a pair of modulated cavities. Our system consists of two identical coupled photonic cavities, often called a ``photonic molecule" \cite{spreeuw_classical_1990, Zhang_2018}, that are modulated to induce transitions between their symmetric and antisymmetric supermodes. This modulation is itself driven by two or more drives at incommensurate frequencies, each of which realizes an orthogonal synthetic dimension. We refer to the lattice sites created by each incommensurate drive as being along a ``Floquet" synthetic dimension. We explicitly construct a 2D Chern insulator that exhibits quantized topological energy pumping, by realizing a driven-dissipative analog of the conservative protocol by Martin, Refael and Halperin \cite{TFC_original}. The rate of energy pumping not only survives the effects of external laser drive but is abetted by the presence of finite dissipation. Moreover, the driven-dissipative nature of our protocol obviates the limitations imposed by finite qubit coherence times and the need for complicated state initialization in other platforms \cite{Boyers_2020, Malz_2021}, attesting to the remarkable topological robustness of the 2D Chern TI. 

Note that our usage of the term Floquet dimensions distinguishes it from synthetic frequency dimensions although both require modulated photonic cavities, as the modulation frequencies in the former case are significantly below the modulation at the free-spectral-range (FSR) required in the latter case. This eases on-chip realization of our approach in integrated photonics by reducing the demanding bandwidths of low-loss electro-optic modulation. The reduction is particularly beneficial for the photonic construction of high-dimensional models as that would require modulation with frequencies 10-100$\times$ the FSR in the synthetic frequency dimension case. 

As an illustration of high-dimensional topology, we use our effectively 0D system to construct a three-dimensional (3D) topological Hamiltonian supporting Weyl points, which act as magnetic monopoles in the reciprocal ($k$) space of the Floquet lattice \cite{Weyl}. To observe this monopole behaviour, we require a reconstruction of the Berry curvature's `field' lines, measured around these Weyl points. We conclude by showing how the Berry curvature can be experimentally measured throughout the bands for any general two-band Hamiltonian.

More generally, harnessing Floquet synthetic dimensions offers a powerful tool for direct $k$-space Hamiltonian engineering in high dimensions using simple, compact geometries with experimentally realizable excitation and measurement protocols.
% \section*{Results}

\subsection*{Half-BHZ model in a photonic molecule and topological energy pumping}

To illustrate the analogous topological behavior of temporally modulated systems in Floquet synthetic dimensions, we consider the Bloch form of the Qi-Wu-Zhang model \cite{qi_topological_2006, che_topological_2020, liang_realization_2023} (equivalently, the half–Bernevig-Hughes-Zhang (half-BHZ) model). When implemented on a Floquet lattice by driving a spin (or a two-level system) with two incommensurate (irrationally related) frequencies \cite{TFC_original}, the Hamiltonian is
\begin{eqnarray}
    \mathcal{H} = \Omega_R[\mathrm{sin}(\Omega_1t + \phi_1)\sigma_x + \mathrm{sin}(\Omega_2t + \phi_2)\sigma_y \nonumber\\
    +\{m - \mathrm{cos}(\Omega_1t + \phi_1) - \mathrm{cos}(\Omega_2t + \phi_2)\}\sigma_z]
\label{eq:BHZ}
\end{eqnarray}
where $\Omega_1t + \phi_1 \rightarrow k_x$ and $\Omega_2t + \phi_2 \rightarrow k_y$ give us the $\vec k$-space representation of the half-BHZ model on a real lattice \cite{bernevig_quantum_2006}. This duality between the drive phases and Bloch quasimomentum $\mathbf{k}$ also implies that the linear evolution of the phase with time emulates the effect of a charge moving under an electric field in a 2D lattice. The half-BHZ Hamiltonian breaks both time-reversal and chiral symmetries but possesses particle-hole symmetry with $\hat C = \sigma_x K$ and inversion symmetry with $\hat P = \sigma_z$, $\hat C \mathcal{H}(k) \hat C^{-1} =-\mathcal{H}(-k) $ and $\hat P \mathcal{H}(k) \hat P^{-1} =\mathcal{H}(-k) $. Thus $\mathcal{H}(\vec k)$ belongs to class D in Altland-Zirnbauer's tenfold way of classification of TIs, and supports chiral transport \cite{schnyder_classification_2008, altland_nonstandard_1997, kitaev_periodic_2009}. It models the behavior of a 2D Chern insulator with a Chern number ($C$) determined by the value of $m$; for $|m|>2$ (trivial phase), $C=0$, and for $|m|<2$ (topological phase), $C=1$. These Chern insulators exhibit an anomalous current that is proportional to $C$, and the current flows perpendicular to the applied field $\vec{\Omega}$. 

Quantized chiral transport is advantageous to probe in Floquet synthetic dimensions as the effective electric field naturally arises here \cite{TFC_original}, while other real-space and synthetic-space systems with neutral particles require effective electric fields to be explicitly introduced. On the Floquet lattice, this leads to topological energy pumping that is quantized by $C$, and we therefore expect to see quantized energy transfer from one incommensurate drive to the other. While implementing this Hamiltonian with qubits \cite{Boyers_2020, Malz_2021} can produce exotic phenomena such as engineering cat states \cite{TFC_Cavity} and quantum state boosting \cite{Long_2022}, they encounter constraints in demonstrating topological pumping on long timescales due to decoherence. However, barring quantum measurements, this Hamiltonian can be simulated by classical systems, such as magnetic nanoparticles in a time-dependent magnetic field, which allows for a demonstration of topological energy pumping that is unimpeded by the coherence times of qubits. We now look at how a photonic molecule driven by two incommensurate frequencies can implement the same physics encapsulated by Eq. \eqref{eq:BHZ}.

A photonic molecule comprises two identical optical ring resonators evanescently coupled at a rate $\mu$. \cite{spreeuw_classical_1990, Zhang_2018}. The eigenmodes of the molecule are the symmetric (S) and anti-symmetric (AS) supermodes of various azimuthal orders as shown in Fig.~\ref{fig:Schematic}. Each ring has an electro-optic phase modulator (EOM) that couples the eigenmodes when driven near the splitting, i.e., $\omega_m\simeq\mu$, with opposite polarities of the RF drive signal $V(t)$. We isolate a single pair of eigenmodes as our %classical 
two-level system, and work in the single-photon subspace. Taking the bosonic annihilation operators for the uncoupled modes of the two rings to be $a_1$ and $a_2$ respectively, we define the S and AS eigenmode annihilation operators as $c_1 = \frac{1}{\sqrt{2}}(a_1 + a_2)$ and $c_2 = \frac{1}{\sqrt{2}}(a_1 - a_2)$, giving us the Hamiltonian,
\begin{equation}
    \mathcal{H} =\omega_+c_1^\dag c_1 + \omega_-c_2^\dag c_2 + gV(t)(c_1^\dag c_2 + c_2^\dag c_1)
\end{equation}
where $g$ is the electro-optic coupling strength. Measuring all frequencies relative to the uncoupled ring resonance frequency $\omega_0$, we set $\omega_\pm = \pm\frac{\mu}{2}$. In this single-photon supermode subspace, the Pauli operators are defined as usual: 
$
    \sigma_x = c_1^\dag c_2 + c_2^\dag c_1, \ 
    i\sigma_y = c_1^\dag c_2 - c_2^\dag c_1, \ 
    \sigma_z = c_1^\dag c_1 - c_2^\dag c_2\ 
    $, 
giving us the Hamiltonian
\begin{equation}
    \mathcal{H} = -\sigma_z\, \mu/2 + \sigma_x \, gV(t)
\end{equation}
We now consider a specific form of $V(t)$ as an RF carrier at $\omega_m$ with I-Q amplitude modulations (AM) $V_x(t)$ and $V_y(t)$ respectively, as well as frequency modulation (FM) $\lambda(t)={\rm d/d}t [\Delta(t)]$, leading to,
\begin{equation*}
    \mathcal{H} = \sigma_z \mu/2 + \sigma_x g  {\rm Re} \left [ \{V_x - iV_y\} \times \exp\{i\omega_m t + i\Delta(t)\}\right]
\end{equation*}
where $\Delta(t) = \int^t\lambda(t')\mathrm{d}t'$. Taking the interaction picture with $|gV(t)|\ll\mu \, \forall\,  t$ (weak driving), the effective Hamiltonian under the rotating-wave approximation is (Supp. I),
\begin{equation}
    \mathcal{H} = -\frac{\delta + \lambda(t)}{2}\sigma_z + \frac{gV_x(t)}{2}\sigma_x + \frac{gV_y(t)}{2}\sigma_y
    \label{eq:SpinH}
\end{equation}
where $\delta = \omega_m - \mu$. Comparing with Eq.~\eqref{eq:BHZ} gives us the necessary amplitude and frequency modulation signals to be applied
\begin{eqnarray}
    \lambda(t) &=& -gV_0\{\mathrm{cos}(\Omega_1t+\phi_1) + \mathrm{cos}(\Omega_2t+\phi_2) \} \nonumber\\
    V_x(t) &=& V_0\,\mathrm{sin}(\Omega_1t+\phi_1)\nonumber\\
    V_y(t) &=& V_0\,\mathrm{sin}(\Omega_2t+\phi_2)
    \label{eq:mod}
\end{eqnarray}
The tunable topological parameter $m=\delta/(2gV_0)$ is now the normalized detuning of the RF drive $\omega_m$ from the resonance of the two-level system and maps to a static $\sigma_z$ coefficient in Eq. \eqref{eq:BHZ}. Thus, $\omega_m$ can be readily controlled from $0<m<2$ to $m>2$ to engender a topological phase transition. The I-Q amplitude modulation maps to $\sigma_x$ and $\sigma_y$ components, whereas the frequency modulation maps to the $\sigma_z$ component.

We note that this system can be made to evolve adiabatically, i.e., $\Omega_1, \Omega_2 \ll gV_0$, by choosing the frequency scales of our modulation accordingly. We also emphasize that all frequency scales in our proposed implementation are much below the ring's free spectral range ($\Omega_1, \Omega_2 \ll \mu \ll \omega_{\rm FSR}$), thus easing bandwidth requirements for integrated photonic modulators. Thus, we have showed how the 2D half-BHZ Hamiltonian can be realized in a photonic molecule, an effectively 0D system, with appropriately engineered drives and detunings, but in an as-of-yet conservative system. Before introducing dissipation and external drive, we briefly summarize the phenomenon of quantized topological energy pumping as introduced in \cite{TFC_original}. Splitting the Hamiltonian into the respective $\Omega_1$ and $\Omega_2$ drive contributions:
\begin{equation}
    \mathcal{H} = h_1(t)+h_2(t) + \sigma_z\, \delta/2,
\end{equation}
the work done by each drive $i=1,2$ over time $T$ is given by
\begin{equation}
    W_i(T) = \int_0^T dt \left\langle\frac{dh_i}{dt}\right\rangle
\end{equation}
For $\Omega_1/\Omega_2$ irrational, the system samples the full Brillouin zone, leading to quantized energy pumping:
\begin{equation}
    W_1 = -W_2 = C\frac{\Omega_1\Omega_2T}{2\pi}
    \label{eq:rate_pumping}
\end{equation}
where $C$ is the Chern number.

\subsection*{Driven-dissipative quantized pumping in a photonic molecule}
\begin{figure*}
    \centering
    \includegraphics[width = 17cm]{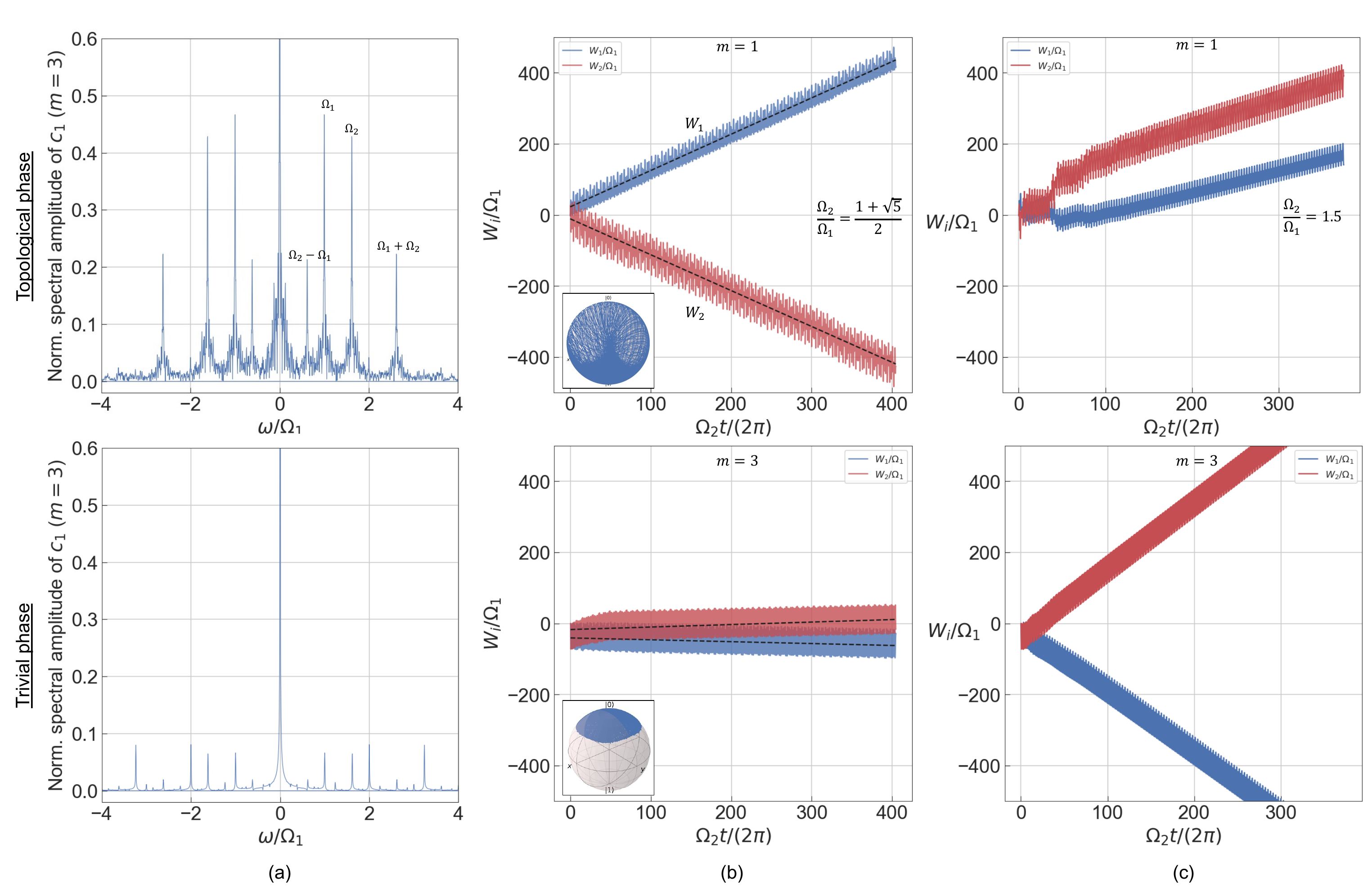}
    \caption{Spectral and temporal signatures of topological dynamics, in the presence of optical drive and dissipation. (a) Normalized spectral amplitude of the symmetric super-mode $c_1$. In the topological regime ($m=1$) we see the dense spectrum showing a continuous floor, indicative of aperiodic and highly delocalized dynamics that are characterized by harmonics of the incommensurate drives, $\Omega_1$ and $\Omega_2$. In the trivial regime ($m=3$), the spectrum is discrete with sharp peaks, indicating periodic orbits and localization in the Bloch sphere. (b) Normalized work done by the drives $\Omega_1$ and $\Omega_2$ in the topological ($m=1$) and trivial ($m=3$) regimes.
    $W_1$ and $W_2$ respectively show slopes of 1.022 and -1.010 in the topological regime over $\sim5$ photon lifetimes, and almost no pumping (slopes of -0.053 and 0.070 respectively for $W_1$ and $W_2$) in the trivial regime, clearly exhibiting the linear dependence on the Chern number $C$. Insets show Bloch sphere trajectories. (c) Dynamics for commensurate drives ($\Omega_2/\Omega_1=1.5$) shows no quantization and the possibility of higher pumping rates in the trivial regime (see text).}
    \label{fig:Results}
\end{figure*}

We next explore how the addition of an external laser drive and cavity dissipation enable quantized topological energy pumping in a quasi-steady state regime without stringent requirements on initialization of the molecule's state.
Dissipation is natural as all optical ring resonators have finite Q-factors, which define the photon decay rate $\gamma$ for the system. Nevertheless, this decay affects both modes symmetrically, and they can be renormalized to look at the dynamics within the two-level subspace. This is a unique advantage that photonic systems provide, and motivates us to verify persistent topological pumping when adding an external optical drive.
The driven-dissipative equations of motion for $c_1$ and $c_2$ are
\begin{equation}
    \dot{c}_{1,2} = i[\mathcal{H},c_{1,2}] - \gamma c_{1,2} + \sqrt{\gamma_e}s_{in}(t)e^{\pm i((\mu+\delta)t+\Delta(t))/2}
\end{equation}
where $\gamma_e$ is the coupling rate into the bus waveguide, and $s_{in}(t)$ is the external laser drive (note that from this point onward in the text, we distinguish between the Hamiltonian drives $\Omega_1$ and $\Omega_2$ and the external optical/laser drive $s_{in}(t)$). 
We physically motivate the chosen laser frequency with the adiabatically varying eigenspectrum of the Hamiltonian in Eq.~\eqref{eq:SpinH}, as a function of $m$. % (Supp. III).
\begin{figure*}
    \centering
    \includegraphics[width = 17cm]{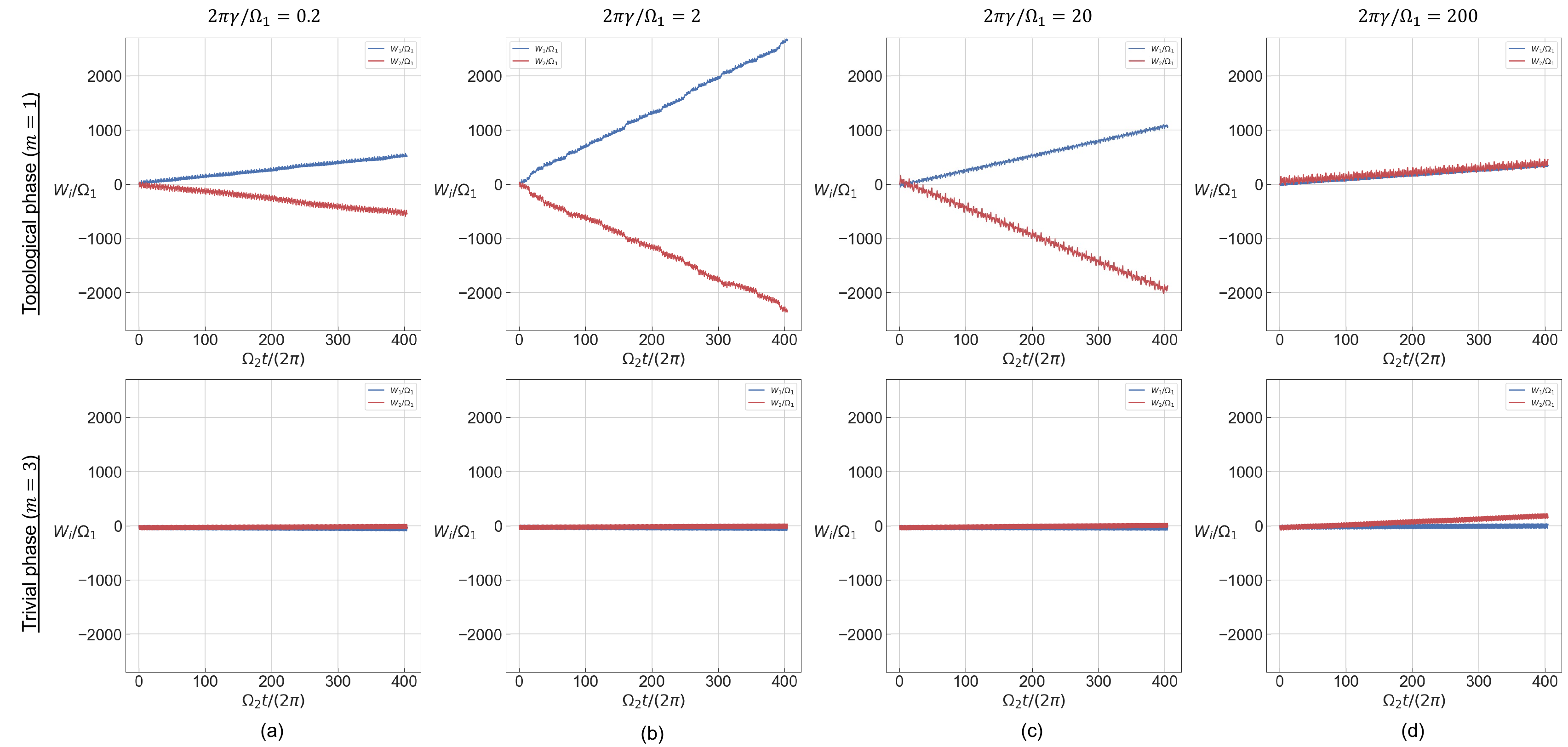}
    \caption{Work done by the drives, simulated for $\gamma = $ (a) $0.1\Omega_1/\pi$, (b) $\Omega_1/\pi$, (c) $10\Omega_1/\pi$, and (d)$100\Omega_1/\pi$. We see close to quantized pumping in (a) with slopes of 1.261 and -1.342 for $W_1$ and $W_2$ respectively, which increases and starts to disappear in (b) and (c). Almost no observable phase transition behavior occurs in (d), due to overdamped oscillator dynamics. The increase in slope is not an anomaly, however, as the actual work done, without normalization depends on $|c_1|^2+|c_2|^2$, which reduces by orders of magnitude and kills the effect of the increased slope (Supp. Fig.~\ref{fig:total_power}) .}
    \label{fig:Dissipation_regimes}
\end{figure*}

\subsection*{Topological signatures in temporal and spectral dynamics}

This system shows a variety of topological signatures even with drive and dissipation added into the mix [Fig.~\ref{fig:Results}], reinforcing the remarkable robust topology of the conservative half-BHZ model. In Fig.~\ref{fig:Results}(a), a clear qualitative difference is seen in the spectral amplitudes of one of the super-modes (denoted by $c_1$): 
The topological regime is characterized by a dense spectrum with a continuous floor, indicative of the aperiodic evolution of the system, while the trivial regime shows several discrete peaks that represents the periodic localization in the dynamics. Our numerical simulations reproduce results calculated in a conservative system averaged over initial states \cite{Crowley_2019}. Note that no such averaging over initial states is required in our model, as the signatures presented are long-time quasi-steady-state simulations; they are fairly independent of the initial state transients which have decayed at long times.

As noted previously, the hallmark of this Hamiltonian is the presence of quantized topological pumping, which is absent in the trivial regime. In Fig.~\ref{fig:Results}(b), we see that the work done by the $\Omega_1$ drive ($W_1$) increases at the quantized rate stated in Eq.~\eqref{eq:rate_pumping}, while the $\Omega_2$ drive ($W_2$) decreases correspondingly, reflecting that there is energy being pumped from one drive to the other. The linear slopes reinforce the quantization by the Chern number $C$, with the pumping rate being 1.022 and -1.010 for $W_1$ and $W_2$ respectively in the topological phase (top figure) and correspondingly -0.053 and 0.070 in the trivial phase (bottom). The Bloch sphere plots (Fig.~\ref{fig:Results}(b) insets, SI video 1 and SI video 2) reflect the ergodic behavior of the system in the topological regime, while the trivial regime dynamics remains localised near the $\ket{0}$ state. Remarkably, we are not limited by the photon lifetime ($\tau_p = 1/\gamma$, with $\gamma \approx 0.01\Omega_1/\pi$), and see sustained pumping for $\sim 5$ photon lifetimes for $m=1$.  Almost no pumping is observed for $m=3$ in the trivial regime. 
An important detail here is that the pumping is being observed under a quasi-steady state condition, due to the drive and dissipation, thus obviating the need for complex Floquet eigenstate initialization in previous protocols \cite{Malz_2021, Boyers_2020} and allowing us to start with vacuum states in both resonators. 

One way to verify the need for incommensurateness in the Hamiltonian's drive frequencies is to observe the dynamics for rational values of $\Omega_1/\Omega_2=p/q, \ p,q \in \mathbb{Z}$. We see the absence of a transition in this case, as the half-BHZ model effectively becomes a 1D tight-binding Hamiltonian with long-range couplings \cite{bell_spectral_2017, dutt_experimental_2019, wang_multidimensional_2020}. The pumping rate no longer depends on the Chern number $C$, but on the integrated Berry curvature over specific periodic trajectories in the Brillouin Zone. We therefore also see periodic orbits on the Bloch sphere (SI video 3 and SI video 4).

\subsection*{Impact of dissipation on topological pumping}
A unique feature of our system is the interplay between the dynamics and the dissipation. We show in Fig.~\ref{fig:Dissipation_regimes} that the topological pumping loses quantization for higher values of $\gamma$, but the striking qualitative contrast between the topological and trivial regimes still persists up to $\gamma > \Omega_1,\Omega_2$. Although the normalized pumping rate seems to increase, the total power $|c_1|^2+|c_2|^2$ reduces on average in the new quasi-steady state, leading to a net reduction in pump power. Physical intuition for the loss of pumping beyond a certain $\gamma\gg 10\Omega_1/\pi$ comes from the coupled oscillator system becoming overdamped, washing out the topological dynamics and the pumping effects.

\subsection*{Scaling to higher dimensions: Weyl points}

\begin{figure}
    \centering
    \includegraphics[width = 8.5cm]{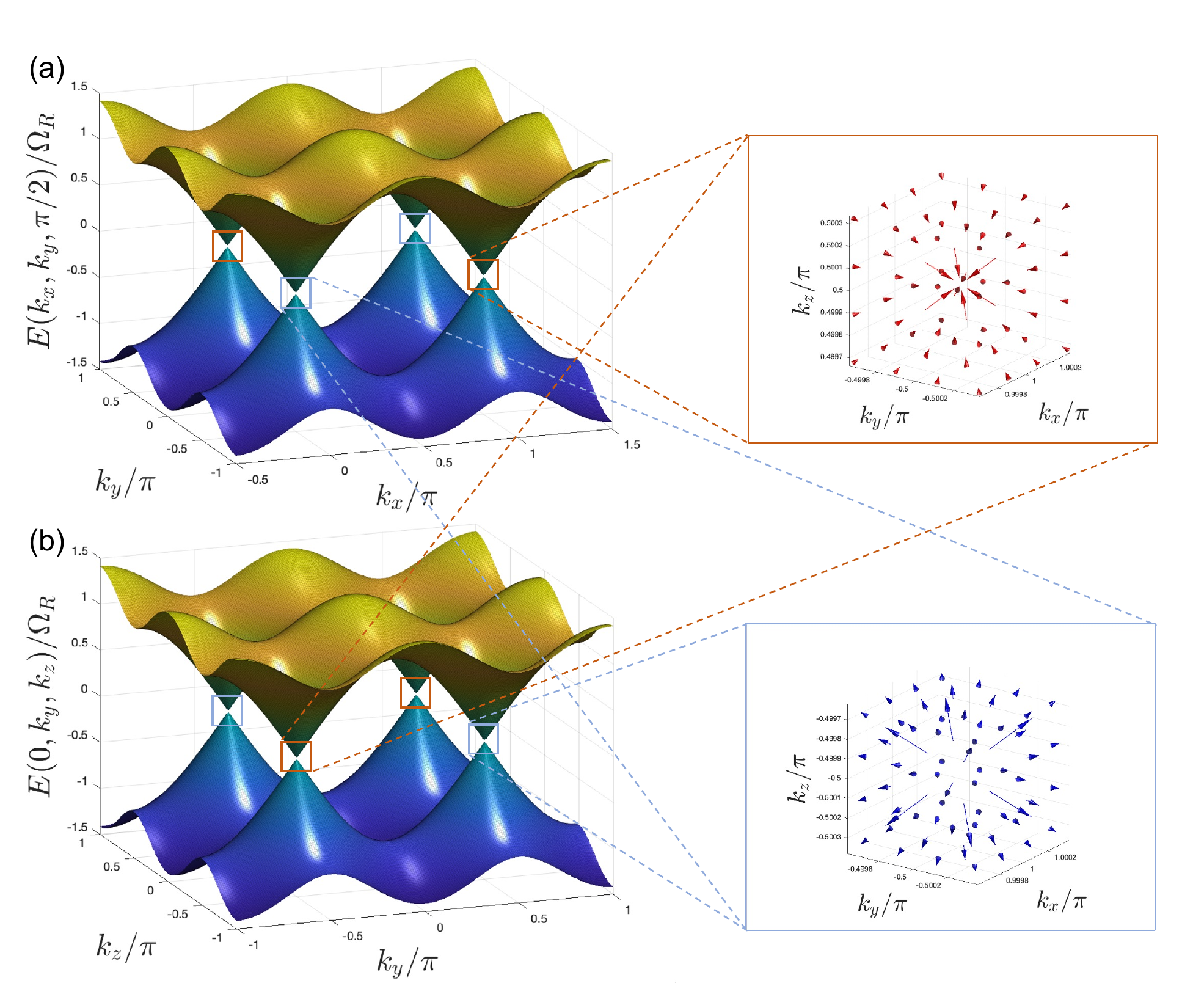}
    \caption{The band structure for a Weyl point Hamiltonian along the (a) $k_xk_y$-plane, and the (b) $k_yk_z$-plane. The band-touching Weyl points show isotropic linear dispersion in their vicinity, with their topological robustness quantified by the Berry curvature flux normal to a surface near the Weyl point being quantized. Neighbouring Weyl points show opposite signs in the quantized flux, indicated by the insets, which show the Berry curvature field lines in 3-D. This behaviour is indicative of synthetic magnetic field lines originating from a Weyl point and converging at the neighbouring Weyl point, leading to the physical picture of Weyl points as monopoles of the synthetic magnetic field.}
    \label{fig:Weyl}
\end{figure}

Now we look at the question of scalability. The key advantage of Floquet synthetic dimensions is the ability to engineer higher-dimensional Hamiltonians, with relatively simple additions to the current setup with a photonic molecule. By simply adding a third incommensurate frequency to the modulation, we can create bulk 3-D non-trivial topology, leading to phenomena such as Weyl points \cite{Weyl}. These points have garnered great interest in the community for the fundamentally unique phenomena they lead to, such as Fermi arc surface states \cite{surfacestates}, which also hold potential for applications in next-generation electronics \cite{expWeyl, Weylelectronics}.

Furthermore, Weyl points behave as monopoles of the Berry curvature (akin to a magnetic field in momentum space), and the Berry curvature surrounding a Weyl point has not been measured experimentally before.
With the photonic molecule, there are two routes to increasing the dimensionality of the system and obtaining Weyl points - introducing a third incommensurate frequency, or harnessing the frequency synthetic dimension of the rings by modulating at the free spectral range \cite{yuan_photonic_2016, dutt_single_2020}.
While modulation at $\omega_{FSR}$ offers the capability to create boundaries on the frequency lattice \cite{dutt2022creating, lu_universal_2020, yuan_photonic_2019, buddhiraju_arbitrary_2021}, adding a third incommensurate frequency gives us the $k$-space Hamiltonian directly. This allows us to simulate Weyl Hamiltonians such as \cite{dubcek_weyl_2015}:
\begin{equation}
    \mathcal{H}(\mathbf{k}) = \Omega_R\{\mathrm{sin}(k_x)\sigma_x+\mathrm{cos}(k_y)\sigma_y+\mathrm{cos}(k_z)\sigma_z\}
    \label{eq:WeylH}
\end{equation}
The band structure of this Hamiltonian is shown in Fig.~\ref{fig:Weyl}, where we see linear dispersion %symmetrically
in all three directions near the band-touching points at ($0,\pm\pi/2,\pm\pi/2$) and ($\pi,\pm\pi/2,\pm\pi/2$).

What makes our method especially powerful is that we can now experimentally map out the Berry curvature throughout the bands, for any given two-band Hamiltonian $\mathcal{H}(\mathbf{k})$, by tracking the dynamics of the artificial spin encoded in the photonic molecule. This allows us to generate Berry curvature data similar to what the insets in Fig.~\ref{fig:Weyl} display in the vicinity of each Weyl point, thus verifying their synthetic monopole character. To buttress this claim, we calculate $\mathbf{B}(\mathbf{k})$ in a gauge-invariant fashion as \cite{bernevig2013topological}:
\begin{widetext}
\begin{equation}
    \mathbf{B}(\mathbf{k}) = \sum_{m \ne n}\mathrm{Im}\left[ \frac{\bra{n(\mathbf{k})}\{\mathbf{\nabla}_\mathbf{k}\mathcal{H}(\mathbf{k})\}\ket{m(\mathbf{k})}\times\bra{m(\mathbf{k})}\{\mathbf{\nabla}_\mathbf{k}\mathcal{H}(\mathbf{k})\}\ket{n(\mathbf{k})}}{(E_m-E_n)^2} \right]
    \label{eq:invariant}
\end{equation}
\end{widetext}
where $\ket{m(\mathbf{k})}$ and $\ket{n(\mathbf{k})}$ are the eigenvectors of $\mathcal{H}(\mathbf{k})$, $E_m$ and $E_n$ being the corresponding eigenvalues. Solving for the general eigenvalues and eigenvectors of the $2\times2$ Hamiltonian (Supp II), this simplifies to
\begin{eqnarray}
    \mathbf{B}(\mathbf{k}) &=\frac{\Omega_R^2}{4E_m^2}\Bigg[
        \hat{k}_x\sin(k_y)\sin(k_z)\left(\bra{n}\sigma_x\ket{n}-\frac{\Omega_R\sin(k_x)}{E_m}\right)\nonumber\\
        &\;\;-\hat{k}_y\cos(k_x)\sin(k_z)\left(\bra{n}\sigma_y\ket{n}-\frac{\Omega_R\cos(k_y)}{E_m}\right)\nonumber\\
        &\;\;-\hat{k}_z\cos(k_x)\sin(k_y)\left(\bra{n}\sigma_z\ket{n}-\frac{\Omega_R\cos(k_z)}{E_m}\right)\Bigg]\nonumber\\
\end{eqnarray}
With Floquet synthetic dimensions in a photonic molecule, we can envision the following protocol: Initialize to the eigenstate $\ket{n}$ for a given initial phase and drive adiabatically with three incommensurate frequencies, while measuring the spin expectation values for all $k$ except at the Weyl points. Here, the eigenstate flips due to the band-touching, and the Berry curvature diverges, so we can threshold this measurement and accumulate experimental runs over multiple initial phases to reconstruct the insets in Fig.~\ref{fig:Weyl}. Thus, the photonic molecule can provide a novel experimental probe for the Berry curvature of a topological Floquet system.

% \section*{Discussion}
\subsection*{Discussion}

In summary, we have proposed a new candidate system to engineer topological Hamiltonians in a 2-D Floquet synthetic lattice, and have shown that it exhibits topological energy pumping that persists in the presence of dissipation and an external drive, over multiple parameter regimes. The pumping persists over multiple photon lifetimes and stays quantized in the normalized spin subspace. Furthermore, this system can be extended to simulate higher-dimensional Hamiltonians with the straightforward addition of an extra incommensurate frequency. As an example, we look at Weyl points and provide a path forward to observing them in a photonic molecule. While synthetic dimensions offer many advantages by circumventing the complexity of constructing analogous material systems, such as in cold-atom experiments \cite{dubcek_weyl_2015}, the complexity inherently shifts to the control signal RF engineering requirements. Furthermore, measuring the topological signatures requires information about the complex amplitudes of $c_1$ and $c_2$, which necessitate dynamically frequency-tuned phase-sensitive detection schemes. However, on-chip electro-optically modulated ring resonators with quality factors upwards of $10^6$, driven by state-of-the-art electronics, have been demonstrated \cite{Zhang_2017, Zhang_2018}. A key advantage lies in the frequencies we modulate at, since $\omega_m\ll\omega_{FSR}$ for Floquet synthetic dimensions, which eases up the specifications on electronics for driving on-chip devices. One can thus envision a fully integrated device that can achieve high-dimensional Hamiltonian analog simulation in this framework for considerable periods of time.

\section*{Acknowledgements}
This work was supported by a Northrop Grumman seed grant and a National Q-Lab joint seed grant from IonQ and the University of Maryland. 

\bibliography{My_Library, mainbib}

\providecommand{\noopsort}[1]{}\providecommand{\singleletter}[1]{#1}%
\begin{thebibliography}{64}%
\makeatletter
\providecommand \@ifxundefined [1]{%
 \@ifx{#1\undefined}
}%
\providecommand \@ifnum [1]{%
 \ifnum #1\expandafter \@firstoftwo
 \else \expandafter \@secondoftwo
 \fi
}%
\providecommand \@ifx [1]{%
 \ifx #1\expandafter \@firstoftwo
 \else \expandafter \@secondoftwo
 \fi
}%
\providecommand \natexlab [1]{#1}%
\providecommand \enquote  [1]{``#1''}%
\providecommand \bibnamefont  [1]{#1}%
\providecommand \bibfnamefont [1]{#1}%
\providecommand \citenamefont [1]{#1}%
\providecommand \href@noop [0]{\@secondoftwo}%
\providecommand \href [0]{\begingroup \@sanitize@url \@href}%
\providecommand \@href[1]{\@@startlink{#1}\@@href}%
\providecommand \@@href[1]{\endgroup#1\@@endlink}%
\providecommand \@sanitize@url [0]{\catcode `\\12\catcode `\$12\catcode
  `\&12\catcode `\#12\catcode `\^12\catcode `\_12\catcode `\%12\relax}%
\providecommand \@@startlink[1]{}%
\providecommand \@@endlink[0]{}%
\providecommand \url  [0]{\begingroup\@sanitize@url \@url }%
\providecommand \@url [1]{\endgroup\@href {#1}{\urlprefix }}%
\providecommand \urlprefix  [0]{URL }%
\providecommand \Eprint [0]{\href }%
\providecommand \doibase [0]{https://doi.org/}%
\providecommand \selectlanguage [0]{\@gobble}%
\providecommand \bibinfo  [0]{\@secondoftwo}%
\providecommand \bibfield  [0]{\@secondoftwo}%
\providecommand \translation [1]{[#1]}%
\providecommand \BibitemOpen [0]{}%
\providecommand \bibitemStop [0]{}%
\providecommand \bibitemNoStop [0]{.\EOS\space}%
\providecommand \EOS [0]{\spacefactor3000\relax}%
\providecommand \BibitemShut  [1]{\csname bibitem#1\endcsname}%
\let\auto@bib@innerbib\@empty
%</preamble>
\bibitem [{\citenamefont {Klitzing}\ \emph
  {et~al.}(1980{\natexlab{a}})\citenamefont {Klitzing}, \citenamefont {Dorda},\
  and\ \citenamefont {Pepper}}]{klitzing_new_1980}%
  \BibitemOpen
  \bibfield  {author} {\bibinfo {author} {\bibfnamefont {K.~v.}\ \bibnamefont
  {Klitzing}}, \bibinfo {author} {\bibfnamefont {G.}~\bibnamefont {Dorda}},\
  and\ \bibinfo {author} {\bibfnamefont {M.}~\bibnamefont {Pepper}},\
  }\bibfield  {title} {\bibinfo {title} {New {Method} for {High}-{Accuracy}
  {Determination} of the {Fine}-{Structure} {Constant} {Based} on {Quantized}
  {Hall} {Resistance}},\ }\href {https://doi.org/10.1103/PhysRevLett.45.494}
  {\bibfield  {journal} {\bibinfo  {journal} {Phys. Rev. Lett.}\ }\textbf
  {\bibinfo {volume} {45}},\ \bibinfo {pages} {494} (\bibinfo {year}
  {1980}{\natexlab{a}})}\BibitemShut {NoStop}%
\bibitem [{\citenamefont {Thouless}\ \emph {et~al.}(1982)\citenamefont
  {Thouless}, \citenamefont {Kohmoto}, \citenamefont {Nightingale},\ and\
  \citenamefont {den Nijs}}]{thouless_quantized_1982}%
  \BibitemOpen
  \bibfield  {author} {\bibinfo {author} {\bibfnamefont {D.~J.}\ \bibnamefont
  {Thouless}}, \bibinfo {author} {\bibfnamefont {M.}~\bibnamefont {Kohmoto}},
  \bibinfo {author} {\bibfnamefont {M.~P.}\ \bibnamefont {Nightingale}},\ and\
  \bibinfo {author} {\bibfnamefont {M.}~\bibnamefont {den Nijs}},\ }\bibfield
  {title} {\bibinfo {title} {Quantized {Hall} {Conductance} in a
  {Two}-{Dimensional} {Periodic} {Potential}},\ }\href
  {https://doi.org/10.1103/PhysRevLett.49.405} {\bibfield  {journal} {\bibinfo
  {journal} {Phys. Rev. Lett.}\ }\textbf {\bibinfo {volume} {49}},\ \bibinfo
  {pages} {405} (\bibinfo {year} {1982})}\BibitemShut {NoStop}%
\bibitem [{\citenamefont {Miyake}\ \emph {et~al.}(2013)\citenamefont {Miyake},
  \citenamefont {Siviloglou}, \citenamefont {Kennedy}, \citenamefont {Burton},\
  and\ \citenamefont {Ketterle}}]{miyake_realizing_2013}%
  \BibitemOpen
  \bibfield  {author} {\bibinfo {author} {\bibfnamefont {H.}~\bibnamefont
  {Miyake}}, \bibinfo {author} {\bibfnamefont {G.~A.}\ \bibnamefont
  {Siviloglou}}, \bibinfo {author} {\bibfnamefont {C.~J.}\ \bibnamefont
  {Kennedy}}, \bibinfo {author} {\bibfnamefont {W.~C.}\ \bibnamefont
  {Burton}},\ and\ \bibinfo {author} {\bibfnamefont {W.}~\bibnamefont
  {Ketterle}},\ }\bibfield  {title} {\bibinfo {title} {Realizing the harper
  hamiltonian with laser-assisted tunneling in optical lattices},\ }\href
  {https://doi.org/10.1103/PhysRevLett.111.185302} {\bibfield  {journal}
  {\bibinfo  {journal} {Phys. Rev. Lett.}\ }\textbf {\bibinfo {volume} {111}},\
  \bibinfo {pages} {185302} (\bibinfo {year} {2013})}\BibitemShut {NoStop}%
\bibitem [{\citenamefont {Aidelsburger}\ \emph {et~al.}(2013)\citenamefont
  {Aidelsburger}, \citenamefont {Atala}, \citenamefont {Lohse}, \citenamefont
  {Barreiro}, \citenamefont {Paredes},\ and\ \citenamefont
  {Bloch}}]{aidelsburger_realization_2013}%
  \BibitemOpen
  \bibfield  {author} {\bibinfo {author} {\bibfnamefont {M.}~\bibnamefont
  {Aidelsburger}}, \bibinfo {author} {\bibfnamefont {M.}~\bibnamefont {Atala}},
  \bibinfo {author} {\bibfnamefont {M.}~\bibnamefont {Lohse}}, \bibinfo
  {author} {\bibfnamefont {J.~T.}\ \bibnamefont {Barreiro}}, \bibinfo {author}
  {\bibfnamefont {B.}~\bibnamefont {Paredes}},\ and\ \bibinfo {author}
  {\bibfnamefont {I.}~\bibnamefont {Bloch}},\ }\bibfield  {title} {\bibinfo
  {title} {Realization of the hofstadter hamiltonian with ultracold atoms in
  optical lattices},\ }\href {https://doi.org/10.1103/PhysRevLett.111.185301}
  {\bibfield  {journal} {\bibinfo  {journal} {Phys. Rev. Lett.}\ }\textbf
  {\bibinfo {volume} {111}},\ \bibinfo {pages} {185301} (\bibinfo {year}
  {2013})}\BibitemShut {NoStop}%
\bibitem [{\citenamefont {Price}\ \emph {et~al.}(2022)\citenamefont {Price},
  \citenamefont {Chong}, \citenamefont {Khanikaev}, \citenamefont {Schomerus},
  \citenamefont {Maczewsky}, \citenamefont {Kremer}, \citenamefont {Heinrich},
  \citenamefont {Szameit}, \citenamefont {Zilberberg}, \citenamefont {Yang},
  \citenamefont {Zhang}, \citenamefont {Alù}, \citenamefont {Thomale},
  \citenamefont {Carusotto}, \citenamefont {St-Jean}, \citenamefont {Amo},
  \citenamefont {Dutt}, \citenamefont {Yuan}, \citenamefont {Fan},
  \citenamefont {Yin}, \citenamefont {Peng}, \citenamefont {Ozawa},\ and\
  \citenamefont {Blanco-Redondo}}]{price_roadmap_2022}%
  \BibitemOpen
  \bibfield  {author} {\bibinfo {author} {\bibfnamefont {H.}~\bibnamefont
  {Price}}, \bibinfo {author} {\bibfnamefont {Y.}~\bibnamefont {Chong}},
  \bibinfo {author} {\bibfnamefont {A.}~\bibnamefont {Khanikaev}}, \bibinfo
  {author} {\bibfnamefont {H.}~\bibnamefont {Schomerus}}, \bibinfo {author}
  {\bibfnamefont {L.~J.}\ \bibnamefont {Maczewsky}}, \bibinfo {author}
  {\bibfnamefont {M.}~\bibnamefont {Kremer}}, \bibinfo {author} {\bibfnamefont
  {M.}~\bibnamefont {Heinrich}}, \bibinfo {author} {\bibfnamefont
  {A.}~\bibnamefont {Szameit}}, \bibinfo {author} {\bibfnamefont
  {O.}~\bibnamefont {Zilberberg}}, \bibinfo {author} {\bibfnamefont
  {Y.}~\bibnamefont {Yang}}, \bibinfo {author} {\bibfnamefont {B.}~\bibnamefont
  {Zhang}}, \bibinfo {author} {\bibfnamefont {A.}~\bibnamefont {Alù}},
  \bibinfo {author} {\bibfnamefont {R.}~\bibnamefont {Thomale}}, \bibinfo
  {author} {\bibfnamefont {I.}~\bibnamefont {Carusotto}}, \bibinfo {author}
  {\bibfnamefont {P.}~\bibnamefont {St-Jean}}, \bibinfo {author} {\bibfnamefont
  {A.}~\bibnamefont {Amo}}, \bibinfo {author} {\bibfnamefont {A.}~\bibnamefont
  {Dutt}}, \bibinfo {author} {\bibfnamefont {L.}~\bibnamefont {Yuan}}, \bibinfo
  {author} {\bibfnamefont {S.}~\bibnamefont {Fan}}, \bibinfo {author}
  {\bibfnamefont {X.}~\bibnamefont {Yin}}, \bibinfo {author} {\bibfnamefont
  {C.}~\bibnamefont {Peng}}, \bibinfo {author} {\bibfnamefont {T.}~\bibnamefont
  {Ozawa}},\ and\ \bibinfo {author} {\bibfnamefont {A.}~\bibnamefont
  {Blanco-Redondo}},\ }\bibfield  {title} {\bibinfo {title} {Roadmap on
  topological photonics},\ }\href {https://doi.org/10.1088/2515-7647/ac4ee4}
  {\bibfield  {journal} {\bibinfo  {journal} {J. Phys. Photonics}\ }\textbf
  {\bibinfo {volume} {4}},\ \bibinfo {pages} {032501} (\bibinfo {year}
  {2022})}\BibitemShut {NoStop}%
\bibitem [{\citenamefont {Khanikaev}\ \emph {et~al.}(2015)\citenamefont
  {Khanikaev}, \citenamefont {Fleury}, \citenamefont {Mousavi},\ and\
  \citenamefont {Alu}}]{khanikaev_topologically_2015}%
  \BibitemOpen
  \bibfield  {author} {\bibinfo {author} {\bibfnamefont {A.~B.}\ \bibnamefont
  {Khanikaev}}, \bibinfo {author} {\bibfnamefont {R.}~\bibnamefont {Fleury}},
  \bibinfo {author} {\bibfnamefont {S.~H.}\ \bibnamefont {Mousavi}},\ and\
  \bibinfo {author} {\bibfnamefont {A.}~\bibnamefont {Alu}},\ }\bibfield
  {title} {\bibinfo {title} {Topologically robust sound propagation in an
  angular-momentum-biased graphene-like resonator lattice},\ }\href@noop {}
  {\bibfield  {journal} {\bibinfo  {journal} {Nature communications}\ }\textbf
  {\bibinfo {volume} {6}},\ \bibinfo {pages} {8260} (\bibinfo {year}
  {2015})}\BibitemShut {NoStop}%
\bibitem [{\citenamefont {Ma}\ \emph {et~al.}(2019)\citenamefont {Ma},
  \citenamefont {Xiao},\ and\ \citenamefont {Chan}}]{ma_topological_2019}%
  \BibitemOpen
  \bibfield  {author} {\bibinfo {author} {\bibfnamefont {G.}~\bibnamefont
  {Ma}}, \bibinfo {author} {\bibfnamefont {M.}~\bibnamefont {Xiao}},\ and\
  \bibinfo {author} {\bibfnamefont {C.~T.}\ \bibnamefont {Chan}},\ }\bibfield
  {title} {\bibinfo {title} {Topological phases in acoustic and mechanical
  systems},\ }\href@noop {} {\bibfield  {journal} {\bibinfo  {journal} {Nature
  Reviews Physics}\ }\textbf {\bibinfo {volume} {1}},\ \bibinfo {pages} {281}
  (\bibinfo {year} {2019})}\BibitemShut {NoStop}%
\bibitem [{\citenamefont {Imhof}\ \emph {et~al.}(2018)\citenamefont {Imhof},
  \citenamefont {Berger}, \citenamefont {Bayer}, \citenamefont {Brehm},
  \citenamefont {Molenkamp}, \citenamefont {Kiessling}, \citenamefont
  {Schindler}, \citenamefont {Lee}, \citenamefont {Greiter}, \citenamefont
  {Neupert},\ and\ \citenamefont {Thomale}}]{imhof_topolectrical-circuit_2018}%
  \BibitemOpen
  \bibfield  {author} {\bibinfo {author} {\bibfnamefont {S.}~\bibnamefont
  {Imhof}}, \bibinfo {author} {\bibfnamefont {C.}~\bibnamefont {Berger}},
  \bibinfo {author} {\bibfnamefont {F.}~\bibnamefont {Bayer}}, \bibinfo
  {author} {\bibfnamefont {J.}~\bibnamefont {Brehm}}, \bibinfo {author}
  {\bibfnamefont {L.~W.}\ \bibnamefont {Molenkamp}}, \bibinfo {author}
  {\bibfnamefont {T.}~\bibnamefont {Kiessling}}, \bibinfo {author}
  {\bibfnamefont {F.}~\bibnamefont {Schindler}}, \bibinfo {author}
  {\bibfnamefont {C.~H.}\ \bibnamefont {Lee}}, \bibinfo {author} {\bibfnamefont
  {M.}~\bibnamefont {Greiter}}, \bibinfo {author} {\bibfnamefont
  {T.}~\bibnamefont {Neupert}},\ and\ \bibinfo {author} {\bibfnamefont
  {R.}~\bibnamefont {Thomale}},\ }\bibfield  {title} {\bibinfo {title}
  {Topolectrical-circuit realization of topological corner modes},\ }\href
  {https://doi.org/10.1038/s41567-018-0246-1} {\bibfield  {journal} {\bibinfo
  {journal} {Nature Physics}\ }\textbf {\bibinfo {volume} {14}},\ \bibinfo
  {pages} {925} (\bibinfo {year} {2018})}\BibitemShut {NoStop}%
\bibitem [{\citenamefont {Hafezi}\ \emph {et~al.}(2013)\citenamefont {Hafezi},
  \citenamefont {Mittal}, \citenamefont {Fan}, \citenamefont {Migdall},\ and\
  \citenamefont {Taylor}}]{hafezi_imaging_2013}%
  \BibitemOpen
  \bibfield  {author} {\bibinfo {author} {\bibfnamefont {M.}~\bibnamefont
  {Hafezi}}, \bibinfo {author} {\bibfnamefont {S.}~\bibnamefont {Mittal}},
  \bibinfo {author} {\bibfnamefont {J.}~\bibnamefont {Fan}}, \bibinfo {author}
  {\bibfnamefont {A.}~\bibnamefont {Migdall}},\ and\ \bibinfo {author}
  {\bibfnamefont {J.~M.}\ \bibnamefont {Taylor}},\ }\bibfield  {title}
  {\bibinfo {title} {Imaging topological edge states in silicon photonics},\
  }\href {https://doi.org/10.1038/nphoton.2013.274} {\bibfield  {journal}
  {\bibinfo  {journal} {Nature Photonics}\ }\textbf {\bibinfo {volume} {7}},\
  \bibinfo {pages} {1001} (\bibinfo {year} {2013})}\BibitemShut {NoStop}%
\bibitem [{\citenamefont {Rechtsman}\ \emph {et~al.}(2013)\citenamefont
  {Rechtsman}, \citenamefont {Zeuner}, \citenamefont {Plotnik}, \citenamefont
  {Lumer}, \citenamefont {Podolsky}, \citenamefont {Dreisow}, \citenamefont
  {Nolte}, \citenamefont {Segev},\ and\ \citenamefont
  {Szameit}}]{rechtsman_photonic_2013}%
  \BibitemOpen
  \bibfield  {author} {\bibinfo {author} {\bibfnamefont {M.~C.}\ \bibnamefont
  {Rechtsman}}, \bibinfo {author} {\bibfnamefont {J.~M.}\ \bibnamefont
  {Zeuner}}, \bibinfo {author} {\bibfnamefont {Y.}~\bibnamefont {Plotnik}},
  \bibinfo {author} {\bibfnamefont {Y.}~\bibnamefont {Lumer}}, \bibinfo
  {author} {\bibfnamefont {D.}~\bibnamefont {Podolsky}}, \bibinfo {author}
  {\bibfnamefont {F.}~\bibnamefont {Dreisow}}, \bibinfo {author} {\bibfnamefont
  {S.}~\bibnamefont {Nolte}}, \bibinfo {author} {\bibfnamefont
  {M.}~\bibnamefont {Segev}},\ and\ \bibinfo {author} {\bibfnamefont
  {A.}~\bibnamefont {Szameit}},\ }\bibfield  {title} {\bibinfo {title}
  {Photonic {Floquet} topological insulators},\ }\href
  {https://doi.org/10.1038/nature12066} {\bibfield  {journal} {\bibinfo
  {journal} {Nature}\ }\textbf {\bibinfo {volume} {496}},\ \bibinfo {pages}
  {196} (\bibinfo {year} {2013})}\BibitemShut {NoStop}%
\bibitem [{\citenamefont {Aidelsburger}\ \emph {et~al.}(2015)\citenamefont
  {Aidelsburger}, \citenamefont {Lohse}, \citenamefont {Schweizer},
  \citenamefont {Atala}, \citenamefont {Barreiro}, \citenamefont {Nascimbène},
  \citenamefont {Cooper}, \citenamefont {Bloch},\ and\ \citenamefont
  {Goldman}}]{aidelsburger_measuring_2015}%
  \BibitemOpen
  \bibfield  {author} {\bibinfo {author} {\bibfnamefont {M.}~\bibnamefont
  {Aidelsburger}}, \bibinfo {author} {\bibfnamefont {M.}~\bibnamefont {Lohse}},
  \bibinfo {author} {\bibfnamefont {C.}~\bibnamefont {Schweizer}}, \bibinfo
  {author} {\bibfnamefont {M.}~\bibnamefont {Atala}}, \bibinfo {author}
  {\bibfnamefont {J.~T.}\ \bibnamefont {Barreiro}}, \bibinfo {author}
  {\bibfnamefont {S.}~\bibnamefont {Nascimbène}}, \bibinfo {author}
  {\bibfnamefont {N.~R.}\ \bibnamefont {Cooper}}, \bibinfo {author}
  {\bibfnamefont {I.}~\bibnamefont {Bloch}},\ and\ \bibinfo {author}
  {\bibfnamefont {N.}~\bibnamefont {Goldman}},\ }\bibfield  {title} {\bibinfo
  {title} {Measuring the {Chern} number of {Hofstadter} bands with ultracold
  bosonic atoms},\ }\href {https://doi.org/10.1038/nphys3171} {\bibfield
  {journal} {\bibinfo  {journal} {Nature Physics}\ }\textbf {\bibinfo {volume}
  {11}},\ \bibinfo {pages} {162} (\bibinfo {year} {2015})}\BibitemShut
  {NoStop}%
\bibitem [{\citenamefont {Wang}\ \emph {et~al.}(2009)\citenamefont {Wang},
  \citenamefont {Chong}, \citenamefont {Joannopoulos},\ and\ \citenamefont
  {Soljačić}}]{wang_observation_2009}%
  \BibitemOpen
  \bibfield  {author} {\bibinfo {author} {\bibfnamefont {Z.}~\bibnamefont
  {Wang}}, \bibinfo {author} {\bibfnamefont {Y.}~\bibnamefont {Chong}},
  \bibinfo {author} {\bibfnamefont {J.~D.}\ \bibnamefont {Joannopoulos}},\ and\
  \bibinfo {author} {\bibfnamefont {M.}~\bibnamefont {Soljačić}},\ }\bibfield
   {title} {\bibinfo {title} {Observation of unidirectional
  backscattering-immune topological electromagnetic states},\ }\href
  {https://doi.org/10.1038/nature08293} {\bibfield  {journal} {\bibinfo
  {journal} {Nature}\ }\textbf {\bibinfo {volume} {461}},\ \bibinfo {pages}
  {772} (\bibinfo {year} {2009})}\BibitemShut {NoStop}%
\bibitem [{\citenamefont {Atala}\ \emph {et~al.}(2014)\citenamefont {Atala},
  \citenamefont {Aidelsburger}, \citenamefont {Lohse}, \citenamefont
  {Barreiro}, \citenamefont {Paredes},\ and\ \citenamefont
  {Bloch}}]{atala_observation_2014}%
  \BibitemOpen
  \bibfield  {author} {\bibinfo {author} {\bibfnamefont {M.}~\bibnamefont
  {Atala}}, \bibinfo {author} {\bibfnamefont {M.}~\bibnamefont {Aidelsburger}},
  \bibinfo {author} {\bibfnamefont {M.}~\bibnamefont {Lohse}}, \bibinfo
  {author} {\bibfnamefont {J.~T.}\ \bibnamefont {Barreiro}}, \bibinfo {author}
  {\bibfnamefont {B.}~\bibnamefont {Paredes}},\ and\ \bibinfo {author}
  {\bibfnamefont {I.}~\bibnamefont {Bloch}},\ }\bibfield  {title} {\bibinfo
  {title} {Observation of chiral currents with ultracold atoms in bosonic
  ladders},\ }\href {https://doi.org/10.1038/nphys2998} {\bibfield  {journal}
  {\bibinfo  {journal} {Nature Physics}\ }\textbf {\bibinfo {volume} {10}},\
  \bibinfo {pages} {588} (\bibinfo {year} {2014})}\BibitemShut {NoStop}%
\bibitem [{\citenamefont {Klitzing}\ \emph
  {et~al.}(1980{\natexlab{b}})\citenamefont {Klitzing}, \citenamefont {Dorda},\
  and\ \citenamefont {Pepper}}]{KvK}%
  \BibitemOpen
  \bibfield  {author} {\bibinfo {author} {\bibfnamefont {K.~v.}\ \bibnamefont
  {Klitzing}}, \bibinfo {author} {\bibfnamefont {G.}~\bibnamefont {Dorda}},\
  and\ \bibinfo {author} {\bibfnamefont {M.}~\bibnamefont {Pepper}},\
  }\bibfield  {title} {\bibinfo {title} {New method for high-accuracy
  determination of the fine-structure constant based on quantized hall
  resistance},\ }\href {https://doi.org/10.1103/PhysRevLett.45.494} {\bibfield
  {journal} {\bibinfo  {journal} {Phys. Rev. Lett.}\ }\textbf {\bibinfo
  {volume} {45}},\ \bibinfo {pages} {494} (\bibinfo {year}
  {1980}{\natexlab{b}})}\BibitemShut {NoStop}%
\bibitem [{\citenamefont {Ozawa}\ and\ \citenamefont
  {Carusotto}(2014)}]{ozawa_anomalous_2014}%
  \BibitemOpen
  \bibfield  {author} {\bibinfo {author} {\bibfnamefont {T.}~\bibnamefont
  {Ozawa}}\ and\ \bibinfo {author} {\bibfnamefont {I.}~\bibnamefont
  {Carusotto}},\ }\bibfield  {title} {\bibinfo {title} {Anomalous and {Quantum}
  {Hall} {Effects} in {Lossy} {Photonic} {Lattices}},\ }\href
  {https://doi.org/10.1103/PhysRevLett.112.133902} {\bibfield  {journal}
  {\bibinfo  {journal} {Phys. Rev. Lett.}\ }\textbf {\bibinfo {volume} {112}},\
  \bibinfo {pages} {133902} (\bibinfo {year} {2014})}\BibitemShut {NoStop}%
\bibitem [{\citenamefont {Yuan}\ \emph {et~al.}(2021)\citenamefont {Yuan},
  \citenamefont {Dutt},\ and\ \citenamefont {Fan}}]{yuan_synthetic_2021}%
  \BibitemOpen
  \bibfield  {author} {\bibinfo {author} {\bibfnamefont {L.}~\bibnamefont
  {Yuan}}, \bibinfo {author} {\bibfnamefont {A.}~\bibnamefont {Dutt}},\ and\
  \bibinfo {author} {\bibfnamefont {S.}~\bibnamefont {Fan}},\ }\bibfield
  {title} {\bibinfo {title} {Synthetic frequency dimensions in dynamically
  modulated ring resonators},\ }\href {https://doi.org/10.1063/5.0056359}
  {\bibfield  {journal} {\bibinfo  {journal} {APL Photonics}\ }\textbf
  {\bibinfo {volume} {6}},\ \bibinfo {pages} {071102} (\bibinfo {year}
  {2021})}\BibitemShut {NoStop}%
\bibitem [{\citenamefont {Ozawa}\ and\ \citenamefont
  {Price}(2019)}]{ozawa_topological_2019}%
  \BibitemOpen
  \bibfield  {author} {\bibinfo {author} {\bibfnamefont {T.}~\bibnamefont
  {Ozawa}}\ and\ \bibinfo {author} {\bibfnamefont {H.~M.}\ \bibnamefont
  {Price}},\ }\bibfield  {title} {\bibinfo {title} {Topological quantum matter
  in synthetic dimensions},\ }\href {https://doi.org/10.1038/s42254-019-0045-3}
  {\bibfield  {journal} {\bibinfo  {journal} {Nature Reviews Physics}\ }\textbf
  {\bibinfo {volume} {1}},\ \bibinfo {pages} {349} (\bibinfo {year}
  {2019})}\BibitemShut {NoStop}%
\bibitem [{\citenamefont {Dutt}\ \emph {et~al.}(2019)\citenamefont {Dutt},
  \citenamefont {Minkov}, \citenamefont {Lin}, \citenamefont {Yuan},
  \citenamefont {Miller},\ and\ \citenamefont {Fan}}]{dutt_experimental_2019}%
  \BibitemOpen
  \bibfield  {author} {\bibinfo {author} {\bibfnamefont {A.}~\bibnamefont
  {Dutt}}, \bibinfo {author} {\bibfnamefont {M.}~\bibnamefont {Minkov}},
  \bibinfo {author} {\bibfnamefont {Q.}~\bibnamefont {Lin}}, \bibinfo {author}
  {\bibfnamefont {L.}~\bibnamefont {Yuan}}, \bibinfo {author} {\bibfnamefont
  {D.~A.~B.}\ \bibnamefont {Miller}},\ and\ \bibinfo {author} {\bibfnamefont
  {S.}~\bibnamefont {Fan}},\ }\bibfield  {title} {\bibinfo {title}
  {Experimental {Demonstration} of {Dynamical} {Input} {Isolation} in
  {Nonadiabatically} {Modulated} {Photonic} {Cavities}},\ }\href
  {https://doi.org/10.1021/acsphotonics.8b01310} {\bibfield  {journal}
  {\bibinfo  {journal} {ACS Photonics}\ }\textbf {\bibinfo {volume} {6}},\
  \bibinfo {pages} {162} (\bibinfo {year} {2019})}\BibitemShut {NoStop}%
\bibitem [{\citenamefont {Leefmans}\ \emph {et~al.}(2022)\citenamefont
  {Leefmans}, \citenamefont {Dutt}, \citenamefont {Williams}, \citenamefont
  {Yuan}, \citenamefont {Parto}, \citenamefont {Nori}, \citenamefont {Fan},\
  and\ \citenamefont {Marandi}}]{leefmans_topological_2022}%
  \BibitemOpen
  \bibfield  {author} {\bibinfo {author} {\bibfnamefont {C.}~\bibnamefont
  {Leefmans}}, \bibinfo {author} {\bibfnamefont {A.}~\bibnamefont {Dutt}},
  \bibinfo {author} {\bibfnamefont {J.}~\bibnamefont {Williams}}, \bibinfo
  {author} {\bibfnamefont {L.}~\bibnamefont {Yuan}}, \bibinfo {author}
  {\bibfnamefont {M.}~\bibnamefont {Parto}}, \bibinfo {author} {\bibfnamefont
  {F.}~\bibnamefont {Nori}}, \bibinfo {author} {\bibfnamefont {S.}~\bibnamefont
  {Fan}},\ and\ \bibinfo {author} {\bibfnamefont {A.}~\bibnamefont {Marandi}},\
  }\bibfield  {title} {\bibinfo {title} {Topological dissipation in a
  time-multiplexed photonic resonator network},\ }\href
  {https://doi.org/10.1038/s41567-021-01492-w} {\bibfield  {journal} {\bibinfo
  {journal} {Nat. Phys.}\ }\textbf {\bibinfo {volume} {18}},\ \bibinfo {pages}
  {442} (\bibinfo {year} {2022})}\BibitemShut {NoStop}%
\bibitem [{\citenamefont {Bartlett}\ \emph {et~al.}(2021)\citenamefont
  {Bartlett}, \citenamefont {Dutt},\ and\ \citenamefont
  {Fan}}]{bartlett_deterministic_2021}%
  \BibitemOpen
  \bibfield  {author} {\bibinfo {author} {\bibfnamefont {B.}~\bibnamefont
  {Bartlett}}, \bibinfo {author} {\bibfnamefont {A.}~\bibnamefont {Dutt}},\
  and\ \bibinfo {author} {\bibfnamefont {S.}~\bibnamefont {Fan}},\ }\bibfield
  {title} {\bibinfo {title} {Deterministic photonic quantum computation in a
  synthetic time dimension},\ }\href {https://doi.org/10.1364/OPTICA.424258}
  {\bibfield  {journal} {\bibinfo  {journal} {Optica}\ }\textbf {\bibinfo
  {volume} {8}},\ \bibinfo {pages} {1515} (\bibinfo {year} {2021})}\BibitemShut
  {NoStop}%
\bibitem [{\citenamefont {Lustig}\ \emph {et~al.}(2019)\citenamefont {Lustig},
  \citenamefont {Weimann}, \citenamefont {Plotnik}, \citenamefont {Lumer},
  \citenamefont {Bandres}, \citenamefont {Szameit},\ and\ \citenamefont
  {Segev}}]{lustig_photonic_2019}%
  \BibitemOpen
  \bibfield  {author} {\bibinfo {author} {\bibfnamefont {E.}~\bibnamefont
  {Lustig}}, \bibinfo {author} {\bibfnamefont {S.}~\bibnamefont {Weimann}},
  \bibinfo {author} {\bibfnamefont {Y.}~\bibnamefont {Plotnik}}, \bibinfo
  {author} {\bibfnamefont {Y.}~\bibnamefont {Lumer}}, \bibinfo {author}
  {\bibfnamefont {M.~A.}\ \bibnamefont {Bandres}}, \bibinfo {author}
  {\bibfnamefont {A.}~\bibnamefont {Szameit}},\ and\ \bibinfo {author}
  {\bibfnamefont {M.}~\bibnamefont {Segev}},\ }\bibfield  {title} {\bibinfo
  {title} {Photonic topological insulator in synthetic dimensions},\ }\href
  {https://doi.org/10.1038/s41586-019-0943-7} {\bibfield  {journal} {\bibinfo
  {journal} {Nature}\ }\textbf {\bibinfo {volume} {567}},\ \bibinfo {pages}
  {356} (\bibinfo {year} {2019})}\BibitemShut {NoStop}%
\bibitem [{\citenamefont {Dutt}\ \emph {et~al.}(2020)\citenamefont {Dutt},
  \citenamefont {Lin}, \citenamefont {Yuan}, \citenamefont {Minkov},
  \citenamefont {Xiao},\ and\ \citenamefont {Fan}}]{dutt_single_2020}%
  \BibitemOpen
  \bibfield  {author} {\bibinfo {author} {\bibfnamefont {A.}~\bibnamefont
  {Dutt}}, \bibinfo {author} {\bibfnamefont {Q.}~\bibnamefont {Lin}}, \bibinfo
  {author} {\bibfnamefont {L.}~\bibnamefont {Yuan}}, \bibinfo {author}
  {\bibfnamefont {M.}~\bibnamefont {Minkov}}, \bibinfo {author} {\bibfnamefont
  {M.}~\bibnamefont {Xiao}},\ and\ \bibinfo {author} {\bibfnamefont
  {S.}~\bibnamefont {Fan}},\ }\bibfield  {title} {\bibinfo {title} {A single
  photonic cavity with two independent physical synthetic dimensions},\ }\href
  {https://doi.org/10.1126/science.aaz3071} {\bibfield  {journal} {\bibinfo
  {journal} {Science}\ }\textbf {\bibinfo {volume} {367}},\ \bibinfo {pages}
  {59} (\bibinfo {year} {2020})}\BibitemShut {NoStop}%
\bibitem [{\citenamefont {Boada}\ \emph {et~al.}(2012)\citenamefont {Boada},
  \citenamefont {Celi}, \citenamefont {Latorre},\ and\ \citenamefont
  {Lewenstein}}]{boada_quantum_2012}%
  \BibitemOpen
  \bibfield  {author} {\bibinfo {author} {\bibfnamefont {O.}~\bibnamefont
  {Boada}}, \bibinfo {author} {\bibfnamefont {A.}~\bibnamefont {Celi}},
  \bibinfo {author} {\bibfnamefont {J.~I.}\ \bibnamefont {Latorre}},\ and\
  \bibinfo {author} {\bibfnamefont {M.}~\bibnamefont {Lewenstein}},\ }\bibfield
   {title} {\bibinfo {title} {Quantum {Simulation} of an {Extra} {Dimension}},\
  }\href {https://doi.org/10.1103/PhysRevLett.108.133001} {\bibfield  {journal}
  {\bibinfo  {journal} {Phys. Rev. Lett.}\ }\textbf {\bibinfo {volume} {108}},\
  \bibinfo {pages} {133001} (\bibinfo {year} {2012})}\BibitemShut {NoStop}%
\bibitem [{\citenamefont {Stuhl}\ \emph {et~al.}(2015)\citenamefont {Stuhl},
  \citenamefont {Lu}, \citenamefont {Aycock}, \citenamefont {Genkina},\ and\
  \citenamefont {Spielman}}]{stuhl_visualizing_2015}%
  \BibitemOpen
  \bibfield  {author} {\bibinfo {author} {\bibfnamefont {B.~K.}\ \bibnamefont
  {Stuhl}}, \bibinfo {author} {\bibfnamefont {H.-I.}\ \bibnamefont {Lu}},
  \bibinfo {author} {\bibfnamefont {L.~M.}\ \bibnamefont {Aycock}}, \bibinfo
  {author} {\bibfnamefont {D.}~\bibnamefont {Genkina}},\ and\ \bibinfo {author}
  {\bibfnamefont {I.~B.}\ \bibnamefont {Spielman}},\ }\bibfield  {title}
  {\bibinfo {title} {Visualizing edge states with an atomic {Bose} gas in the
  quantum {Hall} regime},\ }\href {https://doi.org/10.1126/science.aaa8515}
  {\bibfield  {journal} {\bibinfo  {journal} {Science}\ }\textbf {\bibinfo
  {volume} {349}},\ \bibinfo {pages} {1514} (\bibinfo {year}
  {2015})}\BibitemShut {NoStop}%
\bibitem [{\citenamefont {Mancini}\ \emph {et~al.}(2015)\citenamefont
  {Mancini}, \citenamefont {Pagano}, \citenamefont {Cappellini}, \citenamefont
  {Livi}, \citenamefont {Rider}, \citenamefont {Catani}, \citenamefont {Sias},
  \citenamefont {Zoller}, \citenamefont {Inguscio}, \citenamefont {Dalmonte},\
  and\ \citenamefont {Fallani}}]{mancini_observation_2015}%
  \BibitemOpen
  \bibfield  {author} {\bibinfo {author} {\bibfnamefont {M.}~\bibnamefont
  {Mancini}}, \bibinfo {author} {\bibfnamefont {G.}~\bibnamefont {Pagano}},
  \bibinfo {author} {\bibfnamefont {G.}~\bibnamefont {Cappellini}}, \bibinfo
  {author} {\bibfnamefont {L.}~\bibnamefont {Livi}}, \bibinfo {author}
  {\bibfnamefont {M.}~\bibnamefont {Rider}}, \bibinfo {author} {\bibfnamefont
  {J.}~\bibnamefont {Catani}}, \bibinfo {author} {\bibfnamefont
  {C.}~\bibnamefont {Sias}}, \bibinfo {author} {\bibfnamefont {P.}~\bibnamefont
  {Zoller}}, \bibinfo {author} {\bibfnamefont {M.}~\bibnamefont {Inguscio}},
  \bibinfo {author} {\bibfnamefont {M.}~\bibnamefont {Dalmonte}},\ and\
  \bibinfo {author} {\bibfnamefont {L.}~\bibnamefont {Fallani}},\ }\bibfield
  {title} {\bibinfo {title} {Observation of chiral edge states with neutral
  fermions in synthetic {Hall} ribbons},\ }\href
  {https://doi.org/10.1126/science.aaa8736} {\bibfield  {journal} {\bibinfo
  {journal} {Science}\ }\textbf {\bibinfo {volume} {349}},\ \bibinfo {pages}
  {1510} (\bibinfo {year} {2015})}\BibitemShut {NoStop}%
\bibitem [{\citenamefont {Sundar}\ \emph {et~al.}(2018)\citenamefont {Sundar},
  \citenamefont {Gadway},\ and\ \citenamefont
  {Hazzard}}]{sundar_synthetic_2018}%
  \BibitemOpen
  \bibfield  {author} {\bibinfo {author} {\bibfnamefont {B.}~\bibnamefont
  {Sundar}}, \bibinfo {author} {\bibfnamefont {B.}~\bibnamefont {Gadway}},\
  and\ \bibinfo {author} {\bibfnamefont {K.~R.~A.}\ \bibnamefont {Hazzard}},\
  }\bibfield  {title} {\bibinfo {title} {Synthetic dimensions in ultracold
  polar molecules},\ }\href {https://doi.org/10.1038/s41598-018-21699-x}
  {\bibfield  {journal} {\bibinfo  {journal} {Scientific Reports}\ }\textbf
  {\bibinfo {volume} {8}},\ \bibinfo {pages} {3422} (\bibinfo {year}
  {2018})}\BibitemShut {NoStop}%
\bibitem [{\citenamefont {Luo}\ \emph {et~al.}(2015)\citenamefont {Luo},
  \citenamefont {Zhou}, \citenamefont {Li}, \citenamefont {Xu}, \citenamefont
  {Guo},\ and\ \citenamefont {Zhou}}]{luo_quantum_2015}%
  \BibitemOpen
  \bibfield  {author} {\bibinfo {author} {\bibfnamefont {X.-W.}\ \bibnamefont
  {Luo}}, \bibinfo {author} {\bibfnamefont {X.}~\bibnamefont {Zhou}}, \bibinfo
  {author} {\bibfnamefont {C.-F.}\ \bibnamefont {Li}}, \bibinfo {author}
  {\bibfnamefont {J.-S.}\ \bibnamefont {Xu}}, \bibinfo {author} {\bibfnamefont
  {G.-C.}\ \bibnamefont {Guo}},\ and\ \bibinfo {author} {\bibfnamefont {Z.-W.}\
  \bibnamefont {Zhou}},\ }\bibfield  {title} {\bibinfo {title} {Quantum
  simulation of {2D} topological physics in a {1D} array of optical cavities},\
  }\href {https://doi.org/10.1038/ncomms8704} {\bibfield  {journal} {\bibinfo
  {journal} {Nature Communications}\ }\textbf {\bibinfo {volume} {6}},\
  \bibinfo {pages} {7704} (\bibinfo {year} {2015})}\BibitemShut {NoStop}%
\bibitem [{\citenamefont {Kanungo}\ \emph {et~al.}(2022)\citenamefont
  {Kanungo}, \citenamefont {Whalen}, \citenamefont {Lu}, \citenamefont {Yuan},
  \citenamefont {Dasgupta}, \citenamefont {Dunning}, \citenamefont {Hazzard},\
  and\ \citenamefont {Killian}}]{kanungo_realizing_2022}%
  \BibitemOpen
  \bibfield  {author} {\bibinfo {author} {\bibfnamefont {S.~K.}\ \bibnamefont
  {Kanungo}}, \bibinfo {author} {\bibfnamefont {J.~D.}\ \bibnamefont {Whalen}},
  \bibinfo {author} {\bibfnamefont {Y.}~\bibnamefont {Lu}}, \bibinfo {author}
  {\bibfnamefont {M.}~\bibnamefont {Yuan}}, \bibinfo {author} {\bibfnamefont
  {S.}~\bibnamefont {Dasgupta}}, \bibinfo {author} {\bibfnamefont {F.~B.}\
  \bibnamefont {Dunning}}, \bibinfo {author} {\bibfnamefont {K.~R.~A.}\
  \bibnamefont {Hazzard}},\ and\ \bibinfo {author} {\bibfnamefont {T.~C.}\
  \bibnamefont {Killian}},\ }\bibfield  {title} {\bibinfo {title} {Realizing
  topological edge states with {Rydberg}-atom synthetic dimensions},\ }\href
  {https://doi.org/10.1038/s41467-022-28550-y} {\bibfield  {journal} {\bibinfo
  {journal} {Nat Commun}\ }\textbf {\bibinfo {volume} {13}},\ \bibinfo {pages}
  {972} (\bibinfo {year} {2022})}\BibitemShut {NoStop}%
\bibitem [{\citenamefont {Yang}\ \emph {et~al.}(2023)\citenamefont {Yang},
  \citenamefont {Zhang}, \citenamefont {Liao}, \citenamefont {Liu},
  \citenamefont {Zhou}, \citenamefont {Zhou}, \citenamefont {Xu}, \citenamefont
  {Han}, \citenamefont {Li},\ and\ \citenamefont
  {Guo}}]{yang_realization_2023}%
  \BibitemOpen
  \bibfield  {author} {\bibinfo {author} {\bibfnamefont {M.}~\bibnamefont
  {Yang}}, \bibinfo {author} {\bibfnamefont {H.-Q.}\ \bibnamefont {Zhang}},
  \bibinfo {author} {\bibfnamefont {Y.-W.}\ \bibnamefont {Liao}}, \bibinfo
  {author} {\bibfnamefont {Z.-H.}\ \bibnamefont {Liu}}, \bibinfo {author}
  {\bibfnamefont {Z.-W.}\ \bibnamefont {Zhou}}, \bibinfo {author}
  {\bibfnamefont {X.-X.}\ \bibnamefont {Zhou}}, \bibinfo {author}
  {\bibfnamefont {J.-S.}\ \bibnamefont {Xu}}, \bibinfo {author} {\bibfnamefont
  {Y.-J.}\ \bibnamefont {Han}}, \bibinfo {author} {\bibfnamefont {C.-F.}\
  \bibnamefont {Li}},\ and\ \bibinfo {author} {\bibfnamefont {G.-C.}\
  \bibnamefont {Guo}},\ }\bibfield  {title} {\bibinfo {title} {Realization of
  exceptional points along a synthetic orbital angular momentum dimension},\
  }\href {https://doi.org/10.1126/sciadv.abp8943} {\bibfield  {journal}
  {\bibinfo  {journal} {Science Advances}\ }\textbf {\bibinfo {volume} {9}},\
  \bibinfo {pages} {eabp8943} (\bibinfo {year} {2023})}\BibitemShut {NoStop}%
\bibitem [{\citenamefont {Hu}\ \emph {et~al.}(2020)\citenamefont {Hu},
  \citenamefont {Reimer}, \citenamefont {Shams-Ansari}, \citenamefont {Zhang},\
  and\ \citenamefont {Loncar}}]{hu_realization_2020}%
  \BibitemOpen
  \bibfield  {author} {\bibinfo {author} {\bibfnamefont {Y.}~\bibnamefont
  {Hu}}, \bibinfo {author} {\bibfnamefont {C.}~\bibnamefont {Reimer}}, \bibinfo
  {author} {\bibfnamefont {A.}~\bibnamefont {Shams-Ansari}}, \bibinfo {author}
  {\bibfnamefont {M.}~\bibnamefont {Zhang}},\ and\ \bibinfo {author}
  {\bibfnamefont {M.}~\bibnamefont {Loncar}},\ }\bibfield  {title} {\bibinfo
  {title} {Realization of high-dimensional frequency crystals in electro-optic
  microcombs},\ }\href {https://doi.org/10.1364/OPTICA.395114} {\bibfield
  {journal} {\bibinfo  {journal} {Optica}\ }\textbf {\bibinfo {volume} {7}},\
  \bibinfo {pages} {1189} (\bibinfo {year} {2020})}\BibitemShut {NoStop}%
\bibitem [{\citenamefont {Yuan}\ \emph {et~al.}(2016)\citenamefont {Yuan},
  \citenamefont {Shi},\ and\ \citenamefont {Fan}}]{yuan_photonic_2016}%
  \BibitemOpen
  \bibfield  {author} {\bibinfo {author} {\bibfnamefont {L.}~\bibnamefont
  {Yuan}}, \bibinfo {author} {\bibfnamefont {Y.}~\bibnamefont {Shi}},\ and\
  \bibinfo {author} {\bibfnamefont {S.}~\bibnamefont {Fan}},\ }\bibfield
  {title} {\bibinfo {title} {Photonic gauge potential in a system with a
  synthetic frequency dimension},\ }\href
  {https://doi.org/10.1364/OL.41.000741} {\bibfield  {journal} {\bibinfo
  {journal} {Opt. Lett.}\ }\textbf {\bibinfo {volume} {41}},\ \bibinfo {pages}
  {741} (\bibinfo {year} {2016})}\BibitemShut {NoStop}%
\bibitem [{\citenamefont {Yuan}\ and\ \citenamefont
  {Fan}(2016)}]{yuan_bloch_2016}%
  \BibitemOpen
  \bibfield  {author} {\bibinfo {author} {\bibfnamefont {L.}~\bibnamefont
  {Yuan}}\ and\ \bibinfo {author} {\bibfnamefont {S.}~\bibnamefont {Fan}},\
  }\bibfield  {title} {\bibinfo {title} {Bloch oscillation and unidirectional
  translation of frequency in a dynamically modulated ring resonator},\ }\href
  {https://doi.org/10.1364/OPTICA.3.001014} {\bibfield  {journal} {\bibinfo
  {journal} {Optica}\ }\textbf {\bibinfo {volume} {3}},\ \bibinfo {pages}
  {1014} (\bibinfo {year} {2016})}\BibitemShut {NoStop}%
\bibitem [{\citenamefont {Wang}\ \emph
  {et~al.}(2021{\natexlab{a}})\citenamefont {Wang}, \citenamefont {Dutt},
  \citenamefont {Yang}, \citenamefont {Wojcik}, \citenamefont {Vučković},\
  and\ \citenamefont {Fan}}]{wang_generating_2021}%
  \BibitemOpen
  \bibfield  {author} {\bibinfo {author} {\bibfnamefont {K.}~\bibnamefont
  {Wang}}, \bibinfo {author} {\bibfnamefont {A.}~\bibnamefont {Dutt}}, \bibinfo
  {author} {\bibfnamefont {K.~Y.}\ \bibnamefont {Yang}}, \bibinfo {author}
  {\bibfnamefont {C.~C.}\ \bibnamefont {Wojcik}}, \bibinfo {author}
  {\bibfnamefont {J.}~\bibnamefont {Vučković}},\ and\ \bibinfo {author}
  {\bibfnamefont {S.}~\bibnamefont {Fan}},\ }\bibfield  {title} {\bibinfo
  {title} {Generating arbitrary topological windings of a non-{Hermitian}
  band},\ }\href {https://doi.org/10.1126/science.abf6568} {\bibfield
  {journal} {\bibinfo  {journal} {Science}\ }\textbf {\bibinfo {volume}
  {371}},\ \bibinfo {pages} {1240} (\bibinfo {year}
  {2021}{\natexlab{a}})}\BibitemShut {NoStop}%
\bibitem [{\citenamefont {Wang}\ \emph
  {et~al.}(2021{\natexlab{b}})\citenamefont {Wang}, \citenamefont {Dutt},
  \citenamefont {Wojcik},\ and\ \citenamefont {Fan}}]{wang_topological_2021}%
  \BibitemOpen
  \bibfield  {author} {\bibinfo {author} {\bibfnamefont {K.}~\bibnamefont
  {Wang}}, \bibinfo {author} {\bibfnamefont {A.}~\bibnamefont {Dutt}}, \bibinfo
  {author} {\bibfnamefont {C.~C.}\ \bibnamefont {Wojcik}},\ and\ \bibinfo
  {author} {\bibfnamefont {S.}~\bibnamefont {Fan}},\ }\bibfield  {title}
  {\bibinfo {title} {Topological complex-energy braiding of non-{Hermitian}
  bands},\ }\href {https://doi.org/10.1038/s41586-021-03848-x} {\bibfield
  {journal} {\bibinfo  {journal} {Nature}\ }\textbf {\bibinfo {volume} {598}},\
  \bibinfo {pages} {59} (\bibinfo {year} {2021}{\natexlab{b}})}\BibitemShut
  {NoStop}%
\bibitem [{\citenamefont {Wang}\ \emph {et~al.}(2018)\citenamefont {Wang},
  \citenamefont {Paesani}, \citenamefont {Ding}, \citenamefont {Santagati},
  \citenamefont {Skrzypczyk}, \citenamefont {Salavrakos}, \citenamefont {Tura},
  \citenamefont {Augusiak}, \citenamefont {Mančinska}, \citenamefont {Bacco},
  \citenamefont {Bonneau}, \citenamefont {Silverstone}, \citenamefont {Gong},
  \citenamefont {Acín}, \citenamefont {Rottwitt}, \citenamefont {Oxenløwe},
  \citenamefont {O’Brien}, \citenamefont {Laing},\ and\ \citenamefont
  {Thompson}}]{wang_multidimensional_2018}%
  \BibitemOpen
  \bibfield  {author} {\bibinfo {author} {\bibfnamefont {J.}~\bibnamefont
  {Wang}}, \bibinfo {author} {\bibfnamefont {S.}~\bibnamefont {Paesani}},
  \bibinfo {author} {\bibfnamefont {Y.}~\bibnamefont {Ding}}, \bibinfo {author}
  {\bibfnamefont {R.}~\bibnamefont {Santagati}}, \bibinfo {author}
  {\bibfnamefont {P.}~\bibnamefont {Skrzypczyk}}, \bibinfo {author}
  {\bibfnamefont {A.}~\bibnamefont {Salavrakos}}, \bibinfo {author}
  {\bibfnamefont {J.}~\bibnamefont {Tura}}, \bibinfo {author} {\bibfnamefont
  {R.}~\bibnamefont {Augusiak}}, \bibinfo {author} {\bibfnamefont
  {L.}~\bibnamefont {Mančinska}}, \bibinfo {author} {\bibfnamefont
  {D.}~\bibnamefont {Bacco}}, \bibinfo {author} {\bibfnamefont
  {D.}~\bibnamefont {Bonneau}}, \bibinfo {author} {\bibfnamefont {J.~W.}\
  \bibnamefont {Silverstone}}, \bibinfo {author} {\bibfnamefont
  {Q.}~\bibnamefont {Gong}}, \bibinfo {author} {\bibfnamefont {A.}~\bibnamefont
  {Acín}}, \bibinfo {author} {\bibfnamefont {K.}~\bibnamefont {Rottwitt}},
  \bibinfo {author} {\bibfnamefont {L.~K.}\ \bibnamefont {Oxenløwe}}, \bibinfo
  {author} {\bibfnamefont {J.~L.}\ \bibnamefont {O’Brien}}, \bibinfo {author}
  {\bibfnamefont {A.}~\bibnamefont {Laing}},\ and\ \bibinfo {author}
  {\bibfnamefont {M.~G.}\ \bibnamefont {Thompson}},\ }\bibfield  {title}
  {\bibinfo {title} {Multidimensional quantum entanglement with large-scale
  integrated optics},\ }\href {https://doi.org/10.1126/science.aar7053}
  {\bibfield  {journal} {\bibinfo  {journal} {Science}\ }\textbf {\bibinfo
  {volume} {360}},\ \bibinfo {pages} {285} (\bibinfo {year}
  {2018})}\BibitemShut {NoStop}%
\bibitem [{\citenamefont {Dutt}\ \emph
  {et~al.}(2022{\natexlab{a}})\citenamefont {Dutt}, \citenamefont {Yuan},
  \citenamefont {Yang}, \citenamefont {Wang}, \citenamefont {Buddhiraju},
  \citenamefont {Vučković},\ and\ \citenamefont {Fan}}]{dutt_creating_2022}%
  \BibitemOpen
  \bibfield  {author} {\bibinfo {author} {\bibfnamefont {A.}~\bibnamefont
  {Dutt}}, \bibinfo {author} {\bibfnamefont {L.}~\bibnamefont {Yuan}}, \bibinfo
  {author} {\bibfnamefont {K.~Y.}\ \bibnamefont {Yang}}, \bibinfo {author}
  {\bibfnamefont {K.}~\bibnamefont {Wang}}, \bibinfo {author} {\bibfnamefont
  {S.}~\bibnamefont {Buddhiraju}}, \bibinfo {author} {\bibfnamefont
  {J.}~\bibnamefont {Vučković}},\ and\ \bibinfo {author} {\bibfnamefont
  {S.}~\bibnamefont {Fan}},\ }\bibfield  {title} {\bibinfo {title} {Creating
  boundaries along a synthetic frequency dimension},\ }\href
  {https://doi.org/10.1038/s41467-022-31140-7} {\bibfield  {journal} {\bibinfo
  {journal} {Nat Commun}\ }\textbf {\bibinfo {volume} {13}},\ \bibinfo {pages}
  {3377} (\bibinfo {year} {2022}{\natexlab{a}})}\BibitemShut {NoStop}%
\bibitem [{\citenamefont {Lu}\ \emph {et~al.}(2020)\citenamefont {Lu},
  \citenamefont {Rao}, \citenamefont {Moille}, \citenamefont {Westly},\ and\
  \citenamefont {Srinivasan}}]{lu_universal_2020}%
  \BibitemOpen
  \bibfield  {author} {\bibinfo {author} {\bibfnamefont {X.}~\bibnamefont
  {Lu}}, \bibinfo {author} {\bibfnamefont {A.}~\bibnamefont {Rao}}, \bibinfo
  {author} {\bibfnamefont {G.}~\bibnamefont {Moille}}, \bibinfo {author}
  {\bibfnamefont {D.~A.}\ \bibnamefont {Westly}},\ and\ \bibinfo {author}
  {\bibfnamefont {K.}~\bibnamefont {Srinivasan}},\ }\bibfield  {title}
  {\bibinfo {title} {Universal frequency engineering tool for microcavity
  nonlinear optics: multiple selective mode splitting of whispering-gallery
  resonances},\ }\href {https://doi.org/10.1364/PRJ.401755} {\bibfield
  {journal} {\bibinfo  {journal} {Photon. Res., PRJ}\ }\textbf {\bibinfo
  {volume} {8}},\ \bibinfo {pages} {1676} (\bibinfo {year} {2020})}\BibitemShut
  {NoStop}%
\bibitem [{\citenamefont {Spreeuw}\ \emph {et~al.}(1990)\citenamefont
  {Spreeuw}, \citenamefont {van Druten}, \citenamefont {Beijersbergen},
  \citenamefont {Eliel},\ and\ \citenamefont
  {Woerdman}}]{spreeuw_classical_1990}%
  \BibitemOpen
  \bibfield  {author} {\bibinfo {author} {\bibfnamefont {R.~J.~C.}\
  \bibnamefont {Spreeuw}}, \bibinfo {author} {\bibfnamefont {N.~J.}\
  \bibnamefont {van Druten}}, \bibinfo {author} {\bibfnamefont {M.~W.}\
  \bibnamefont {Beijersbergen}}, \bibinfo {author} {\bibfnamefont {E.~R.}\
  \bibnamefont {Eliel}},\ and\ \bibinfo {author} {\bibfnamefont {J.~P.}\
  \bibnamefont {Woerdman}},\ }\bibfield  {title} {\bibinfo {title} {Classical
  realization of a strongly driven two-level system},\ }\href
  {https://doi.org/10.1103/PhysRevLett.65.2642} {\bibfield  {journal} {\bibinfo
   {journal} {Phys. Rev. Lett.}\ }\textbf {\bibinfo {volume} {65}},\ \bibinfo
  {pages} {2642} (\bibinfo {year} {1990})}\BibitemShut {NoStop}%
\bibitem [{\citenamefont {Zhang}\ \emph {et~al.}(2018)\citenamefont {Zhang},
  \citenamefont {Wang}, \citenamefont {Hu}, \citenamefont {Shams-Ansari},
  \citenamefont {Ren}, \citenamefont {Fan},\ and\ \citenamefont
  {Lon{\v{c}}ar}}]{Zhang_2018}%
  \BibitemOpen
  \bibfield  {author} {\bibinfo {author} {\bibfnamefont {M.}~\bibnamefont
  {Zhang}}, \bibinfo {author} {\bibfnamefont {C.}~\bibnamefont {Wang}},
  \bibinfo {author} {\bibfnamefont {Y.}~\bibnamefont {Hu}}, \bibinfo {author}
  {\bibfnamefont {A.}~\bibnamefont {Shams-Ansari}}, \bibinfo {author}
  {\bibfnamefont {T.}~\bibnamefont {Ren}}, \bibinfo {author} {\bibfnamefont
  {S.}~\bibnamefont {Fan}},\ and\ \bibinfo {author} {\bibfnamefont
  {M.}~\bibnamefont {Lon{\v{c}}ar}},\ }\bibfield  {title} {\bibinfo {title}
  {Electronically programmable photonic molecule},\ }\href
  {https://doi.org/10.1038/s41566-018-0317-y} {\bibfield  {journal} {\bibinfo
  {journal} {Nature Photonics}\ }\textbf {\bibinfo {volume} {13}},\ \bibinfo
  {pages} {36} (\bibinfo {year} {2018})}\BibitemShut {NoStop}%
\bibitem [{\citenamefont {Martin}\ \emph {et~al.}(2017)\citenamefont {Martin},
  \citenamefont {Refael},\ and\ \citenamefont {Halperin}}]{TFC_original}%
  \BibitemOpen
  \bibfield  {author} {\bibinfo {author} {\bibfnamefont {I.}~\bibnamefont
  {Martin}}, \bibinfo {author} {\bibfnamefont {G.}~\bibnamefont {Refael}},\
  and\ \bibinfo {author} {\bibfnamefont {B.}~\bibnamefont {Halperin}},\
  }\bibfield  {title} {\bibinfo {title} {Topological frequency conversion in
  strongly driven quantum systems},\ }\href
  {https://doi.org/10.1103/PhysRevX.7.041008} {\bibfield  {journal} {\bibinfo
  {journal} {Phys. Rev. X}\ }\textbf {\bibinfo {volume} {7}},\ \bibinfo {pages}
  {041008} (\bibinfo {year} {2017})}\BibitemShut {NoStop}%
\bibitem [{\citenamefont {Boyers}\ \emph {et~al.}(2020)\citenamefont {Boyers},
  \citenamefont {Crowley}, \citenamefont {Chandran},\ and\ \citenamefont
  {Sushkov}}]{Boyers_2020}%
  \BibitemOpen
  \bibfield  {author} {\bibinfo {author} {\bibfnamefont {E.}~\bibnamefont
  {Boyers}}, \bibinfo {author} {\bibfnamefont {P.~J.~D.}\ \bibnamefont
  {Crowley}}, \bibinfo {author} {\bibfnamefont {A.}~\bibnamefont {Chandran}},\
  and\ \bibinfo {author} {\bibfnamefont {A.~O.}\ \bibnamefont {Sushkov}},\
  }\bibfield  {title} {\bibinfo {title} {Exploring 2d synthetic quantum hall
  physics with a quasiperiodically driven qubit},\ }\href
  {https://doi.org/10.1103/PhysRevLett.125.160505} {\bibfield  {journal}
  {\bibinfo  {journal} {Phys. Rev. Lett.}\ }\textbf {\bibinfo {volume} {125}},\
  \bibinfo {pages} {160505} (\bibinfo {year} {2020})}\BibitemShut {NoStop}%
\bibitem [{\citenamefont {Malz}\ and\ \citenamefont {Smith}(2021)}]{Malz_2021}%
  \BibitemOpen
  \bibfield  {author} {\bibinfo {author} {\bibfnamefont {D.}~\bibnamefont
  {Malz}}\ and\ \bibinfo {author} {\bibfnamefont {A.}~\bibnamefont {Smith}},\
  }\bibfield  {title} {\bibinfo {title} {Topological two-dimensional floquet
  lattice on a single superconducting qubit},\ }\href
  {https://doi.org/10.1103/PhysRevLett.126.163602} {\bibfield  {journal}
  {\bibinfo  {journal} {Phys. Rev. Lett.}\ }\textbf {\bibinfo {volume} {126}},\
  \bibinfo {pages} {163602} (\bibinfo {year} {2021})}\BibitemShut {NoStop}%
\bibitem [{\citenamefont {Weyl}(1929)}]{Weyl}%
  \BibitemOpen
  \bibfield  {author} {\bibinfo {author} {\bibfnamefont {H.}~\bibnamefont
  {Weyl}},\ }\href {https://doi.org/10.1007/BF01339504} {\bibfield  {journal}
  {\bibinfo  {journal} {Z. Physik}\ }\textbf {\bibinfo {volume} {56}},\
  \bibinfo {pages} {330} (\bibinfo {year} {1929})}\BibitemShut {NoStop}%
\bibitem [{\citenamefont {Qi}\ \emph {et~al.}(2006)\citenamefont {Qi},
  \citenamefont {Wu},\ and\ \citenamefont {Zhang}}]{qi_topological_2006}%
  \BibitemOpen
  \bibfield  {author} {\bibinfo {author} {\bibfnamefont {X.-L.}\ \bibnamefont
  {Qi}}, \bibinfo {author} {\bibfnamefont {Y.-S.}\ \bibnamefont {Wu}},\ and\
  \bibinfo {author} {\bibfnamefont {S.-C.}\ \bibnamefont {Zhang}},\ }\bibfield
  {title} {\bibinfo {title} {Topological quantization of the spin {Hall} effect
  in two-dimensional paramagnetic semiconductors},\ }\href
  {https://doi.org/10.1103/PhysRevB.74.085308} {\bibfield  {journal} {\bibinfo
  {journal} {Phys. Rev. B}\ }\textbf {\bibinfo {volume} {74}},\ \bibinfo
  {pages} {085308} (\bibinfo {year} {2006})}\BibitemShut {NoStop}%
\bibitem [{\citenamefont {Che}\ \emph {et~al.}(2020)\citenamefont {Che},
  \citenamefont {Gneiting}, \citenamefont {Liu},\ and\ \citenamefont
  {Nori}}]{che_topological_2020}%
  \BibitemOpen
  \bibfield  {author} {\bibinfo {author} {\bibfnamefont {Y.}~\bibnamefont
  {Che}}, \bibinfo {author} {\bibfnamefont {C.}~\bibnamefont {Gneiting}},
  \bibinfo {author} {\bibfnamefont {T.}~\bibnamefont {Liu}},\ and\ \bibinfo
  {author} {\bibfnamefont {F.}~\bibnamefont {Nori}},\ }\bibfield  {title}
  {\bibinfo {title} {Topological quantum phase transitions retrieved through
  unsupervised machine learning},\ }\href
  {https://doi.org/10.1103/PhysRevB.102.134213} {\bibfield  {journal} {\bibinfo
   {journal} {Phys. Rev. B}\ }\textbf {\bibinfo {volume} {102}},\ \bibinfo
  {pages} {134213} (\bibinfo {year} {2020})}\BibitemShut {NoStop}%
\bibitem [{\citenamefont {Liang}\ \emph {et~al.}(2023)\citenamefont {Liang},
  \citenamefont {Wei}, \citenamefont {Zhang}, \citenamefont {Wang},
  \citenamefont {Zhang}, \citenamefont {Wang}, \citenamefont {Qi},
  \citenamefont {Liu},\ and\ \citenamefont {Zhang}}]{liang_realization_2023}%
  \BibitemOpen
  \bibfield  {author} {\bibinfo {author} {\bibfnamefont {M.-C.}\ \bibnamefont
  {Liang}}, \bibinfo {author} {\bibfnamefont {Y.-D.}\ \bibnamefont {Wei}},
  \bibinfo {author} {\bibfnamefont {L.}~\bibnamefont {Zhang}}, \bibinfo
  {author} {\bibfnamefont {X.-J.}\ \bibnamefont {Wang}}, \bibinfo {author}
  {\bibfnamefont {H.}~\bibnamefont {Zhang}}, \bibinfo {author} {\bibfnamefont
  {W.-W.}\ \bibnamefont {Wang}}, \bibinfo {author} {\bibfnamefont
  {W.}~\bibnamefont {Qi}}, \bibinfo {author} {\bibfnamefont {X.-J.}\
  \bibnamefont {Liu}},\ and\ \bibinfo {author} {\bibfnamefont {X.}~\bibnamefont
  {Zhang}},\ }\bibfield  {title} {\bibinfo {title} {Realization of
  {Qi}-{Wu}-{Zhang} model in spin-orbit-coupled ultracold fermions},\ }\href
  {https://doi.org/10.1103/PhysRevResearch.5.L012006} {\bibfield  {journal}
  {\bibinfo  {journal} {Phys. Rev. Res.}\ }\textbf {\bibinfo {volume} {5}},\
  \bibinfo {pages} {L012006} (\bibinfo {year} {2023})}\BibitemShut {NoStop}%
\bibitem [{\citenamefont {Bernevig}\ \emph {et~al.}(2006)\citenamefont
  {Bernevig}, \citenamefont {Hughes},\ and\ \citenamefont
  {Zhang}}]{bernevig_quantum_2006}%
  \BibitemOpen
  \bibfield  {author} {\bibinfo {author} {\bibfnamefont {B.~A.}\ \bibnamefont
  {Bernevig}}, \bibinfo {author} {\bibfnamefont {T.~L.}\ \bibnamefont
  {Hughes}},\ and\ \bibinfo {author} {\bibfnamefont {S.-C.}\ \bibnamefont
  {Zhang}},\ }\bibfield  {title} {\bibinfo {title} {Quantum {Spin} {Hall}
  {Effect} and {Topological} {Phase} {Transition} in {HgTe} {Quantum}
  {Wells}},\ }\href {https://doi.org/10.1126/science.1133734} {\bibfield
  {journal} {\bibinfo  {journal} {Science}\ }\textbf {\bibinfo {volume}
  {314}},\ \bibinfo {pages} {1757} (\bibinfo {year} {2006})}\BibitemShut
  {NoStop}%
\bibitem [{\citenamefont {Schnyder}\ \emph {et~al.}(2008)\citenamefont
  {Schnyder}, \citenamefont {Ryu}, \citenamefont {Furusaki},\ and\
  \citenamefont {Ludwig}}]{schnyder_classification_2008}%
  \BibitemOpen
  \bibfield  {author} {\bibinfo {author} {\bibfnamefont {A.~P.}\ \bibnamefont
  {Schnyder}}, \bibinfo {author} {\bibfnamefont {S.}~\bibnamefont {Ryu}},
  \bibinfo {author} {\bibfnamefont {A.}~\bibnamefont {Furusaki}},\ and\
  \bibinfo {author} {\bibfnamefont {A.~W.~W.}\ \bibnamefont {Ludwig}},\
  }\bibfield  {title} {\bibinfo {title} {Classification of topological
  insulators and superconductors in three spatial dimensions},\ }\href
  {https://doi.org/10.1103/PhysRevB.78.195125} {\bibfield  {journal} {\bibinfo
  {journal} {Phys. Rev. B}\ }\textbf {\bibinfo {volume} {78}},\ \bibinfo
  {pages} {195125} (\bibinfo {year} {2008})}\BibitemShut {NoStop}%
\bibitem [{\citenamefont {Altland}\ and\ \citenamefont
  {Zirnbauer}(1997)}]{altland_nonstandard_1997}%
  \BibitemOpen
  \bibfield  {author} {\bibinfo {author} {\bibfnamefont {A.}~\bibnamefont
  {Altland}}\ and\ \bibinfo {author} {\bibfnamefont {M.~R.}\ \bibnamefont
  {Zirnbauer}},\ }\bibfield  {title} {\bibinfo {title} {Nonstandard symmetry
  classes in mesoscopic normal-superconducting hybrid structures},\ }\href
  {https://doi.org/10.1103/PhysRevB.55.1142} {\bibfield  {journal} {\bibinfo
  {journal} {Phys. Rev. B}\ }\textbf {\bibinfo {volume} {55}},\ \bibinfo
  {pages} {1142} (\bibinfo {year} {1997})}\BibitemShut {NoStop}%
\bibitem [{\citenamefont {Kitaev}(2009)}]{kitaev_periodic_2009}%
  \BibitemOpen
  \bibfield  {author} {\bibinfo {author} {\bibfnamefont {A.}~\bibnamefont
  {Kitaev}},\ }\bibfield  {title} {\bibinfo {title} {Periodic table for
  topological insulators and superconductors},\ }\href
  {https://doi.org/10.1063/1.3149495} {\bibfield  {journal} {\bibinfo
  {journal} {AIP Conference Proceedings}\ }\textbf {\bibinfo {volume} {1134}},\
  \bibinfo {pages} {22} (\bibinfo {year} {2009})}\BibitemShut {NoStop}%
\bibitem [{\citenamefont {Nathan}\ \emph {et~al.}(2019)\citenamefont {Nathan},
  \citenamefont {Martin},\ and\ \citenamefont {Refael}}]{TFC_Cavity}%
  \BibitemOpen
  \bibfield  {author} {\bibinfo {author} {\bibfnamefont {F.}~\bibnamefont
  {Nathan}}, \bibinfo {author} {\bibfnamefont {I.}~\bibnamefont {Martin}},\
  and\ \bibinfo {author} {\bibfnamefont {G.}~\bibnamefont {Refael}},\
  }\bibfield  {title} {\bibinfo {title} {Topological frequency conversion in a
  driven dissipative quantum cavity},\ }\href
  {https://doi.org/10.1103/PhysRevB.99.094311} {\bibfield  {journal} {\bibinfo
  {journal} {Phys. Rev. B}\ }\textbf {\bibinfo {volume} {99}},\ \bibinfo
  {pages} {094311} (\bibinfo {year} {2019})}\BibitemShut {NoStop}%
\bibitem [{\citenamefont {Long}\ \emph {et~al.}(2022)\citenamefont {Long},
  \citenamefont {Crowley}, \citenamefont {Koll{\'{a}}r},\ and\ \citenamefont
  {Chandran}}]{Long_2022}%
  \BibitemOpen
  \bibfield  {author} {\bibinfo {author} {\bibfnamefont {D.~M.}\ \bibnamefont
  {Long}}, \bibinfo {author} {\bibfnamefont {P.~J.}\ \bibnamefont {Crowley}},
  \bibinfo {author} {\bibfnamefont {A.~J.}\ \bibnamefont {Koll{\'{a}}r}},\ and\
  \bibinfo {author} {\bibfnamefont {A.}~\bibnamefont {Chandran}},\ }\bibfield
  {title} {\bibinfo {title} {Boosting the quantum state of a cavity with
  floquet driving},\ }\href {https://doi.org/10.1103/physrevlett.128.183602}
  {\bibfield  {journal} {\bibinfo  {journal} {Physical Review Letters}\
  }\textbf {\bibinfo {volume} {128}},\ \bibinfo {pages} {183602} (\bibinfo
  {year} {2022})}\BibitemShut {NoStop}%
\bibitem [{\citenamefont {Crowley}\ \emph {et~al.}(2019)\citenamefont
  {Crowley}, \citenamefont {Martin},\ and\ \citenamefont
  {Chandran}}]{Crowley_2019}%
  \BibitemOpen
  \bibfield  {author} {\bibinfo {author} {\bibfnamefont {P.~J.~D.}\
  \bibnamefont {Crowley}}, \bibinfo {author} {\bibfnamefont {I.}~\bibnamefont
  {Martin}},\ and\ \bibinfo {author} {\bibfnamefont {A.}~\bibnamefont
  {Chandran}},\ }\bibfield  {title} {\bibinfo {title} {Topological
  classification of quasiperiodically driven quantum systems},\ }\bibfield
  {journal} {\bibinfo  {journal} {Physical Review B}\ }\textbf {\bibinfo
  {volume} {99}},\ \href {https://doi.org/10.1103/physrevb.99.064306}
  {10.1103/physrevb.99.064306} (\bibinfo {year} {2019})\BibitemShut {NoStop}%
\bibitem [{\citenamefont {Bell}\ \emph {et~al.}(2017)\citenamefont {Bell},
  \citenamefont {Wang}, \citenamefont {Solntsev}, \citenamefont {Neshev},
  \citenamefont {Sukhorukov},\ and\ \citenamefont
  {Eggleton}}]{bell_spectral_2017}%
  \BibitemOpen
  \bibfield  {author} {\bibinfo {author} {\bibfnamefont {B.~A.}\ \bibnamefont
  {Bell}}, \bibinfo {author} {\bibfnamefont {K.}~\bibnamefont {Wang}}, \bibinfo
  {author} {\bibfnamefont {A.~S.}\ \bibnamefont {Solntsev}}, \bibinfo {author}
  {\bibfnamefont {D.~N.}\ \bibnamefont {Neshev}}, \bibinfo {author}
  {\bibfnamefont {A.~A.}\ \bibnamefont {Sukhorukov}},\ and\ \bibinfo {author}
  {\bibfnamefont {B.~J.}\ \bibnamefont {Eggleton}},\ }\bibfield  {title}
  {\bibinfo {title} {Spectral photonic lattices with complex long-range
  coupling},\ }\href {https://doi.org/10.1364/OPTICA.4.001433} {\bibfield
  {journal} {\bibinfo  {journal} {Optica}\ }\textbf {\bibinfo {volume} {4}},\
  \bibinfo {pages} {1433} (\bibinfo {year} {2017})}\BibitemShut {NoStop}%
\bibitem [{\citenamefont {Wang}\ \emph {et~al.}(2020)\citenamefont {Wang},
  \citenamefont {Bell}, \citenamefont {Solntsev}, \citenamefont {Neshev},
  \citenamefont {Eggleton},\ and\ \citenamefont
  {Sukhorukov}}]{wang_multidimensional_2020}%
  \BibitemOpen
  \bibfield  {author} {\bibinfo {author} {\bibfnamefont {K.}~\bibnamefont
  {Wang}}, \bibinfo {author} {\bibfnamefont {B.~A.}\ \bibnamefont {Bell}},
  \bibinfo {author} {\bibfnamefont {A.~S.}\ \bibnamefont {Solntsev}}, \bibinfo
  {author} {\bibfnamefont {D.~N.}\ \bibnamefont {Neshev}}, \bibinfo {author}
  {\bibfnamefont {B.~J.}\ \bibnamefont {Eggleton}},\ and\ \bibinfo {author}
  {\bibfnamefont {A.~A.}\ \bibnamefont {Sukhorukov}},\ }\bibfield  {title}
  {\bibinfo {title} {Multidimensional synthetic chiral-tube lattices via
  nonlinear frequency conversion},\ }\href
  {https://doi.org/10.1038/s41377-020-0299-7} {\bibfield  {journal} {\bibinfo
  {journal} {Light: Science \& Applications}\ }\textbf {\bibinfo {volume}
  {9}},\ \bibinfo {pages} {132} (\bibinfo {year} {2020})}\BibitemShut {NoStop}%
\bibitem [{\citenamefont {Turner}\ \emph {et~al.}(2013)\citenamefont {Turner},
  \citenamefont {Vishwanath},\ and\ \citenamefont {Head}}]{surfacestates}%
  \BibitemOpen
  \bibfield  {author} {\bibinfo {author} {\bibfnamefont {A.~M.}\ \bibnamefont
  {Turner}}, \bibinfo {author} {\bibfnamefont {A.}~\bibnamefont {Vishwanath}},\
  and\ \bibinfo {author} {\bibfnamefont {C.~O.}\ \bibnamefont {Head}},\
  }\bibfield  {title} {\bibinfo {title} {Beyond band insulators: topology of
  semimetals and interacting phases},\ }\href@noop {} {\bibfield  {journal}
  {\bibinfo  {journal} {Topological Insulators}\ }\textbf {\bibinfo {volume}
  {6}},\ \bibinfo {pages} {293} (\bibinfo {year} {2013})}\BibitemShut {NoStop}%
\bibitem [{\citenamefont {Xu}\ \emph {et~al.}(2015)\citenamefont {Xu},
  \citenamefont {Belopolski}, \citenamefont {Alidoust}, \citenamefont
  {Neupane}, \citenamefont {Bian}, \citenamefont {Zhang}, \citenamefont
  {Sankar}, \citenamefont {Chang}, \citenamefont {Yuan}, \citenamefont {Lee}
  \emph {et~al.}}]{expWeyl}%
  \BibitemOpen
  \bibfield  {author} {\bibinfo {author} {\bibfnamefont {S.-Y.}\ \bibnamefont
  {Xu}}, \bibinfo {author} {\bibfnamefont {I.}~\bibnamefont {Belopolski}},
  \bibinfo {author} {\bibfnamefont {N.}~\bibnamefont {Alidoust}}, \bibinfo
  {author} {\bibfnamefont {M.}~\bibnamefont {Neupane}}, \bibinfo {author}
  {\bibfnamefont {G.}~\bibnamefont {Bian}}, \bibinfo {author} {\bibfnamefont
  {C.}~\bibnamefont {Zhang}}, \bibinfo {author} {\bibfnamefont
  {R.}~\bibnamefont {Sankar}}, \bibinfo {author} {\bibfnamefont
  {G.}~\bibnamefont {Chang}}, \bibinfo {author} {\bibfnamefont
  {Z.}~\bibnamefont {Yuan}}, \bibinfo {author} {\bibfnamefont {C.-C.}\
  \bibnamefont {Lee}}, \emph {et~al.},\ }\bibfield  {title} {\bibinfo {title}
  {Discovery of a weyl fermion semimetal and topological fermi arcs},\
  }\href@noop {} {\bibfield  {journal} {\bibinfo  {journal} {Science}\ }\textbf
  {\bibinfo {volume} {349}},\ \bibinfo {pages} {613} (\bibinfo {year}
  {2015})}\BibitemShut {NoStop}%
\bibitem [{\citenamefont {Osterhoudt}\ \emph {et~al.}(2019)\citenamefont
  {Osterhoudt}, \citenamefont {Diebel}, \citenamefont {Gray}, \citenamefont
  {Yang}, \citenamefont {Stanco}, \citenamefont {Huang}, \citenamefont {Shen},
  \citenamefont {Ni}, \citenamefont {Moll}, \citenamefont {Ran} \emph
  {et~al.}}]{Weylelectronics}%
  \BibitemOpen
  \bibfield  {author} {\bibinfo {author} {\bibfnamefont {G.~B.}\ \bibnamefont
  {Osterhoudt}}, \bibinfo {author} {\bibfnamefont {L.~K.}\ \bibnamefont
  {Diebel}}, \bibinfo {author} {\bibfnamefont {M.~J.}\ \bibnamefont {Gray}},
  \bibinfo {author} {\bibfnamefont {X.}~\bibnamefont {Yang}}, \bibinfo {author}
  {\bibfnamefont {J.}~\bibnamefont {Stanco}}, \bibinfo {author} {\bibfnamefont
  {X.}~\bibnamefont {Huang}}, \bibinfo {author} {\bibfnamefont
  {B.}~\bibnamefont {Shen}}, \bibinfo {author} {\bibfnamefont {N.}~\bibnamefont
  {Ni}}, \bibinfo {author} {\bibfnamefont {P.~J.}\ \bibnamefont {Moll}},
  \bibinfo {author} {\bibfnamefont {Y.}~\bibnamefont {Ran}}, \emph {et~al.},\
  }\bibfield  {title} {\bibinfo {title} {Colossal mid-infrared bulk
  photovoltaic effect in a type-i weyl semimetal},\ }\href@noop {} {\bibfield
  {journal} {\bibinfo  {journal} {Nature materials}\ }\textbf {\bibinfo
  {volume} {18}},\ \bibinfo {pages} {471} (\bibinfo {year} {2019})}\BibitemShut
  {NoStop}%
\bibitem [{\citenamefont {Dutt}\ \emph
  {et~al.}(2022{\natexlab{b}})\citenamefont {Dutt}, \citenamefont {Yuan},
  \citenamefont {Yang}, \citenamefont {Wang}, \citenamefont {Buddhiraju},
  \citenamefont {Vu{\v{c}}kovi{\'c}},\ and\ \citenamefont
  {Fan}}]{dutt2022creating}%
  \BibitemOpen
  \bibfield  {author} {\bibinfo {author} {\bibfnamefont {A.}~\bibnamefont
  {Dutt}}, \bibinfo {author} {\bibfnamefont {L.}~\bibnamefont {Yuan}}, \bibinfo
  {author} {\bibfnamefont {K.~Y.}\ \bibnamefont {Yang}}, \bibinfo {author}
  {\bibfnamefont {K.}~\bibnamefont {Wang}}, \bibinfo {author} {\bibfnamefont
  {S.}~\bibnamefont {Buddhiraju}}, \bibinfo {author} {\bibfnamefont
  {J.}~\bibnamefont {Vu{\v{c}}kovi{\'c}}},\ and\ \bibinfo {author}
  {\bibfnamefont {S.}~\bibnamefont {Fan}},\ }\bibfield  {title} {\bibinfo
  {title} {Creating boundaries along a synthetic frequency dimension},\
  }\href@noop {} {\bibfield  {journal} {\bibinfo  {journal} {Nature
  Communications}\ }\textbf {\bibinfo {volume} {13}},\ \bibinfo {pages} {3377}
  (\bibinfo {year} {2022}{\natexlab{b}})}\BibitemShut {NoStop}%
\bibitem [{\citenamefont {Yuan}\ \emph {et~al.}(2019)\citenamefont {Yuan},
  \citenamefont {Lin}, \citenamefont {Zhang}, \citenamefont {Xiao},
  \citenamefont {Chen},\ and\ \citenamefont {Fan}}]{yuan_photonic_2019}%
  \BibitemOpen
  \bibfield  {author} {\bibinfo {author} {\bibfnamefont {L.}~\bibnamefont
  {Yuan}}, \bibinfo {author} {\bibfnamefont {Q.}~\bibnamefont {Lin}}, \bibinfo
  {author} {\bibfnamefont {A.}~\bibnamefont {Zhang}}, \bibinfo {author}
  {\bibfnamefont {M.}~\bibnamefont {Xiao}}, \bibinfo {author} {\bibfnamefont
  {X.}~\bibnamefont {Chen}},\ and\ \bibinfo {author} {\bibfnamefont
  {S.}~\bibnamefont {Fan}},\ }\bibfield  {title} {\bibinfo {title} {Photonic
  {Gauge} {Potential} in {One} {Cavity} with {Synthetic} {Frequency} and
  {Orbital} {Angular} {Momentum} {Dimensions}},\ }\href
  {https://doi.org/10.1103/PhysRevLett.122.083903} {\bibfield  {journal}
  {\bibinfo  {journal} {Physical Review Letters}\ }\textbf {\bibinfo {volume}
  {122}},\ \bibinfo {pages} {083903} (\bibinfo {year} {2019})}\BibitemShut
  {NoStop}%
\bibitem [{\citenamefont {Buddhiraju}\ \emph {et~al.}(2021)\citenamefont
  {Buddhiraju}, \citenamefont {Dutt}, \citenamefont {Minkov}, \citenamefont
  {Williamson},\ and\ \citenamefont {Fan}}]{buddhiraju_arbitrary_2021}%
  \BibitemOpen
  \bibfield  {author} {\bibinfo {author} {\bibfnamefont {S.}~\bibnamefont
  {Buddhiraju}}, \bibinfo {author} {\bibfnamefont {A.}~\bibnamefont {Dutt}},
  \bibinfo {author} {\bibfnamefont {M.}~\bibnamefont {Minkov}}, \bibinfo
  {author} {\bibfnamefont {I.~A.~D.}\ \bibnamefont {Williamson}},\ and\
  \bibinfo {author} {\bibfnamefont {S.}~\bibnamefont {Fan}},\ }\bibfield
  {title} {\bibinfo {title} {Arbitrary linear transformations for photons in
  the frequency synthetic dimension},\ }\href
  {https://doi.org/10.1038/s41467-021-22670-7} {\bibfield  {journal} {\bibinfo
  {journal} {Nature Communications}\ }\textbf {\bibinfo {volume} {12}},\
  \bibinfo {pages} {2401} (\bibinfo {year} {2021})}\BibitemShut {NoStop}%
\bibitem [{\citenamefont {Dubček}\ \emph {et~al.}(2015)\citenamefont
  {Dubček}, \citenamefont {Kennedy}, \citenamefont {Lu}, \citenamefont
  {Ketterle}, \citenamefont {Soljačić},\ and\ \citenamefont
  {Buljan}}]{dubcek_weyl_2015}%
  \BibitemOpen
  \bibfield  {author} {\bibinfo {author} {\bibfnamefont {T.}~\bibnamefont
  {Dubček}}, \bibinfo {author} {\bibfnamefont {C.~J.}\ \bibnamefont
  {Kennedy}}, \bibinfo {author} {\bibfnamefont {L.}~\bibnamefont {Lu}},
  \bibinfo {author} {\bibfnamefont {W.}~\bibnamefont {Ketterle}}, \bibinfo
  {author} {\bibfnamefont {M.}~\bibnamefont {Soljačić}},\ and\ \bibinfo
  {author} {\bibfnamefont {H.}~\bibnamefont {Buljan}},\ }\bibfield  {title}
  {\bibinfo {title} {Weyl {Points} in {Three}-{Dimensional} {Optical}
  {Lattices}: {Synthetic} {Magnetic} {Monopoles} in {Momentum} {Space}},\
  }\href {https://doi.org/10.1103/PhysRevLett.114.225301} {\bibfield  {journal}
  {\bibinfo  {journal} {Phys. Rev. Lett.}\ }\textbf {\bibinfo {volume} {114}},\
  \bibinfo {pages} {225301} (\bibinfo {year} {2015})}\BibitemShut {NoStop}%
\bibitem [{\citenamefont {Bernevig}(2013)}]{bernevig2013topological}%
  \BibitemOpen
  \bibfield  {author} {\bibinfo {author} {\bibfnamefont {B.~A.}\ \bibnamefont
  {Bernevig}},\ }\bibfield  {title} {\bibinfo {title} {Topological insulators
  and topological superconductors},\ }in\ \href@noop {} {\emph {\bibinfo
  {booktitle} {Topological Insulators and Topological Superconductors}}}\
  (\bibinfo  {publisher} {Princeton university press},\ \bibinfo {year}
  {2013})\BibitemShut {NoStop}%
\bibitem [{\citenamefont {Zhang}\ \emph {et~al.}(2017)\citenamefont {Zhang},
  \citenamefont {Wang}, \citenamefont {Cheng}, \citenamefont {Shams-Ansari},\
  and\ \citenamefont {Lon{\v{c} }ar}}]{Zhang_2017}%
  \BibitemOpen
  \bibfield  {author} {\bibinfo {author} {\bibfnamefont {M.}~\bibnamefont
  {Zhang}}, \bibinfo {author} {\bibfnamefont {C.}~\bibnamefont {Wang}},
  \bibinfo {author} {\bibfnamefont {R.}~\bibnamefont {Cheng}}, \bibinfo
  {author} {\bibfnamefont {A.}~\bibnamefont {Shams-Ansari}},\ and\ \bibinfo
  {author} {\bibfnamefont {M.}~\bibnamefont {Lon{\v{c} }ar}},\ }\bibfield
  {title} {\bibinfo {title} {Monolithic ultra-high-q lithium niobate microring
  resonator},\ }\href {https://doi.org/10.1364/optica.4.001536} {\bibfield
  {journal} {\bibinfo  {journal} {Optica}\ }\textbf {\bibinfo {volume} {4}},\
  \bibinfo {pages} {1536} (\bibinfo {year} {2017})}\BibitemShut {NoStop}%
\end{thebibliography}%
% Produces the bibliography via BibTeX.

\clearpage
\widetext
%%%%%%%%%% Merge with supplemental materials %%%%%%%%%%
%%%%%%%%%% Prefix a "S" to all equations, figures, tables and reset the counter %%%%%%%%%%
\setcounter{equation}{0}
\setcounter{figure}{0}
\setcounter{table}{0}
\setcounter{page}{1}
\makeatletter
\renewcommand{\theequation}{S\arabic{equation}}
\renewcommand{\thefigure}{S\arabic{figure}}
\renewcommand{\bibnumfmt}[1]{[S#1]}
\renewcommand{\citenumfont}[1]{S#1}

\begin{center}
\textbf{\large Supplemental Materials: Quantized topological energy pumping and Weyl points in\\Floquet synthetic dimensions with a driven-dissipative photonic molecule}
\end{center}

    \section{Half-BHZ Hamiltonian for a Photonic Molecule}
    Consider the two-resonator system with a Hamiltonian
    \begin{align}
        H = \omega_0 \left( a_1^\dagger a_1 + a_2^\dagger a_2 \right) + \frac{\mu}{2} \left( a_1^\dagger a_2 + a_2^\dagger a_1 \right) + gV(t)(a_1^\dagger a_1 - a_2 a_2^\dagger)
    \end{align}
    where $a_1$ and $a_2$ are the respective bosonic annihilation operators for the individual ring resonator modes at the resonance $\omega_0$, with a coupling rate $\mu$, and an electro-optic coupling coefficient $g$. Upon diagonalizing the time-independent terms, the Hamiltonian becomes
    \begin{align}
        H = \omega_+ c_1^\dagger c_1 + \omega_- c_2^\dagger c_2 + gV(t)(c_1^\dag c_2 + c_2^\dag c_1)
    \end{align}
    with $\displaystyle{\omega_\pm = \omega_0 \pm \frac{\mu}{2}}$ and $\displaystyle{c_1 = \frac{1}{\sqrt 2} \left( a_1 + a_2 \right), c_2 = \frac{1}{\sqrt 2} \left( a_1 - a_2 \right)}$. With a variable change, we can rotate out the $\omega_0 \left( c_1^\dagger c_1 + c_2^\dagger c_2 \right)$ term (within the single photon subspace) and consider $\omega_\pm = \pm \dfrac{\mu}{2}$. We specify $V(t)$ to have a carrier tone $\omega_m$ with independent amplitude I-Q modulations and frequency modulation, giving us
    \begin{align}
        H &= \omega_+ c_1^\dagger c_1 + \omega_- c_2^\dagger c_2 + g\left[ V_x (t) \cos \left( \omega_m t + \Delta(t) \right) + V_y (t) \sin \left( \omega_m t + \Delta(t) \right) \right]\,( c_1^\dagger c_2 + c_2^\dagger c_1)
    \end{align}
    We can define our spin-matrices to be 
    \begin{align*}
        \sigma_x &= \left( c_1^\dagger c_2 + c_2^\dagger c_1 \right) \\
        \sigma_y &= -i\left( c_1^\dagger c_2 - c_2^\dagger c_1\right) \\
        \sigma_z &= \left( c_1^\dagger c_1 - c_2^\dagger c_2 \right)
    \end{align*}
    in the single-photon subspace. These operators now form a Pauli group and obey the same commutation relations. In terms of these new spin operators, the Hamiltonian is
    \begin{gather}
        H = \frac{\mu}{2} \sigma_z + gV(t) \sigma_x = \frac{\mu}{2} \sigma_z + g\left[ V_x (t) \cos \left\{ \omega_m t + \Delta(t) \right\} + V_y (t) \sin \left\{ \omega_m t + \Delta(t) \right\} \right] \sigma_x
        \label{h1}
    \end{gather}
    In the interaction picture, we can consider $U = \exp\left[i\sigma_z\{\omega_m t + \Delta(t)\}/2\right]$, and the dynamic Hamiltonian becomes
    \begin{gather}
        V_I(t) = \exp\left[i\sigma_z\{\omega_m t + \Delta(t)\}/2\right]\,\left(gV(t)\sigma_x\right) \exp\left[-i\sigma_z\{\omega_m t + \Delta(t)\}/2\right] = gV(t) \sum_{k = 0}^\infty \frac{(i\omega_m t + \Delta(t))^k}{2^k k!} \left[ \sigma_z, \sigma_x \right]_{(k)}
    \end{gather}
    where
    \begin{align*}
        \left[ A, B \right]_{0} &= B \\
        \left[ A, B \right]_{1} &= [A, B] \\
        \left[ A, B \right]_{2} &= [A, [A, B]] \\
        \left[ A, B \right]_{3} &= [A, [A, [A, B]]]
    \end{align*}
    and so on. Since $[\sigma_z, \sigma_x] = 2i\sigma_y, [\sigma_z, \sigma_y] = -2i\sigma_x$,
    \begin{gather}
        V_I (t) = gV (t) \left[- \sum_{k} \frac{(-1)^k (\omega_m t + \Delta(t))^{(2k+1)}}{(2k+1)!} \sigma_y + \sum_{k} \frac{(-1)^k (\omega_m t + \Delta(t))^{2k}}{(2k)!} \sigma_x \right] \nonumber \\ = gV(t) \left[ -\sin \{ \omega_m t + \Delta(t)\} \sigma_y + \cos \{\omega_m t + \Delta(t)\} \sigma_x \right] \nonumber \\
        = g \left[ V_x (t) \cos \left\{ \omega_m t + \Delta(t) \right\} + V_y (t) \sin \left\{ \omega_m t + \Delta(t) \right\} \right] \left[ -\sin \{ \omega_m t + \Delta(t)\} \sigma_y + \cos \{\omega_m t + \Delta(t)\} \sigma_x \right]
    \end{gather}
   For $|gV_x(t)|,|gV_y(t)|\ll\omega_m$, we can expand and apply the rotating wave approximation to $V_I$, keeping only the slowly rotating terms.
    \begin{align}
        V_I (t) &= g\left( e^{i\left( \omega_m t + \Delta(t) \right)} \left( V_x(t) - iV_y(t) \right) + c.c. \right)  \left( e^{i(\omega_m t + \Delta(t))} (\sigma_x+i\sigma_y)/2 + h.c. \right) \nonumber \\
        &\approx \frac{gV_x(t)}{2} \sigma_x + \frac{gV_y(t)}{2} \sigma_y
    \end{align}
    In this rotated frame, the Hamiltonian becomes,
    \begin{align}
        \mathcal H &= UHU^\dagger + i\frac{\partial U}{\partial t}U^\dagger \nonumber\\
        &=-\frac{\delta + \lambda (t)}{2} \sigma_z + \frac{gV_x(t)}{2} \sigma_x + \frac{gV_y(t)}{2} \sigma_y \label{h2}
    \end{align}

    \section{Weyl points in Floquet Synthetic Dimensions}
    %Put math to show that we can get Weyl points with 3 incommensurate frequencies - calculate Chern numbers & Berry curvatures for the Weyl points in Suppl., derive band-structure from k-space Hamiltonian diagonalization
    \begin{figure}
        \centering
        \includegraphics[width = 17cm]{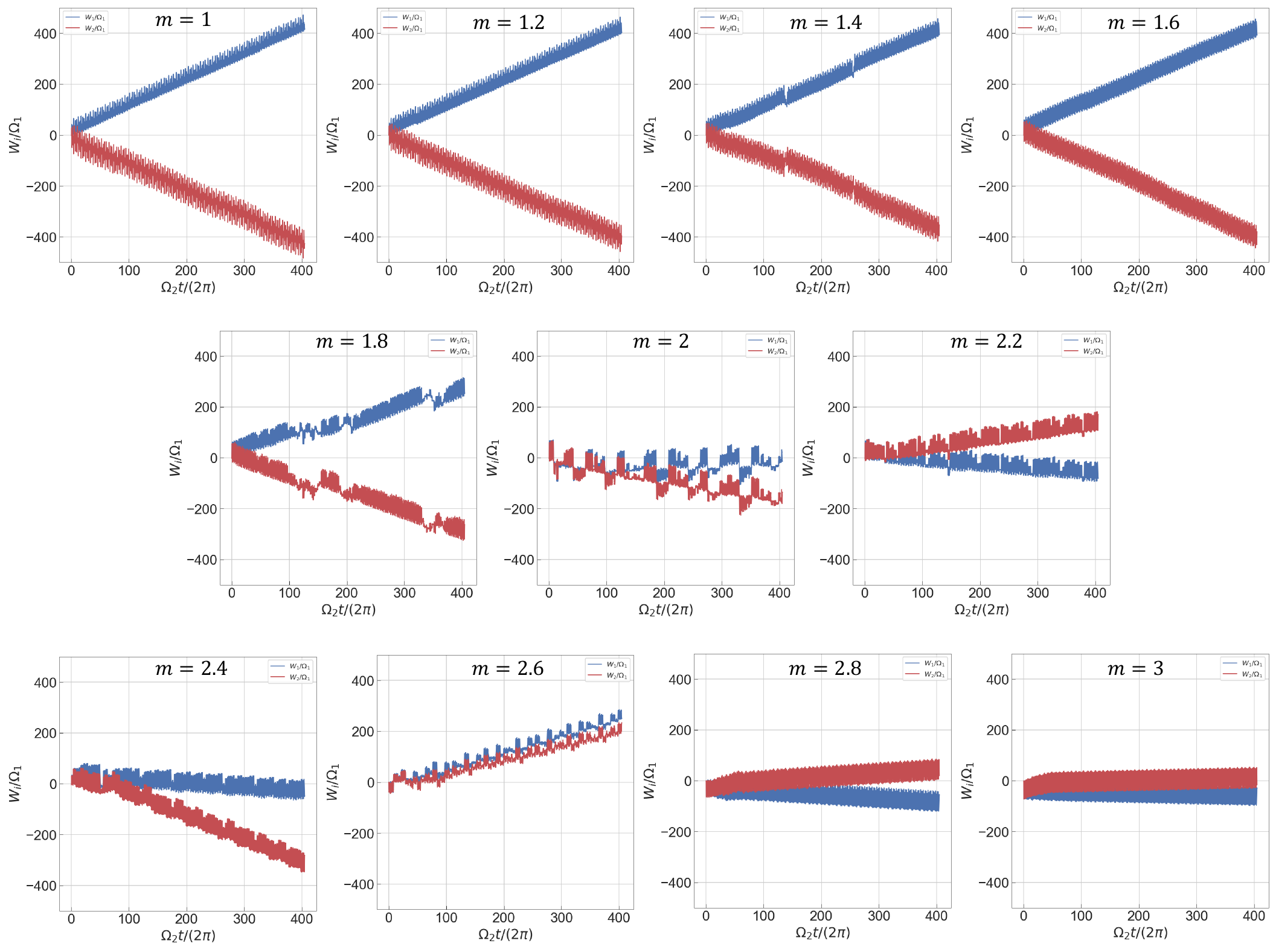}     
        \caption{Topological energy pumping for $\Omega_2/\Omega_1 = (1+\sqrt{5})/2$. We see a qualitative change in behavior around $m=2$, indicative of the topological transition. We simulate this with $gV_0 = 40\Omega_1$ and $\gamma = 0.01\Omega_1/\pi$.}
        \label{fig:golden_sweep}
    \end{figure}
    \begin{figure}
        \centering
        \includegraphics[width = 17cm]{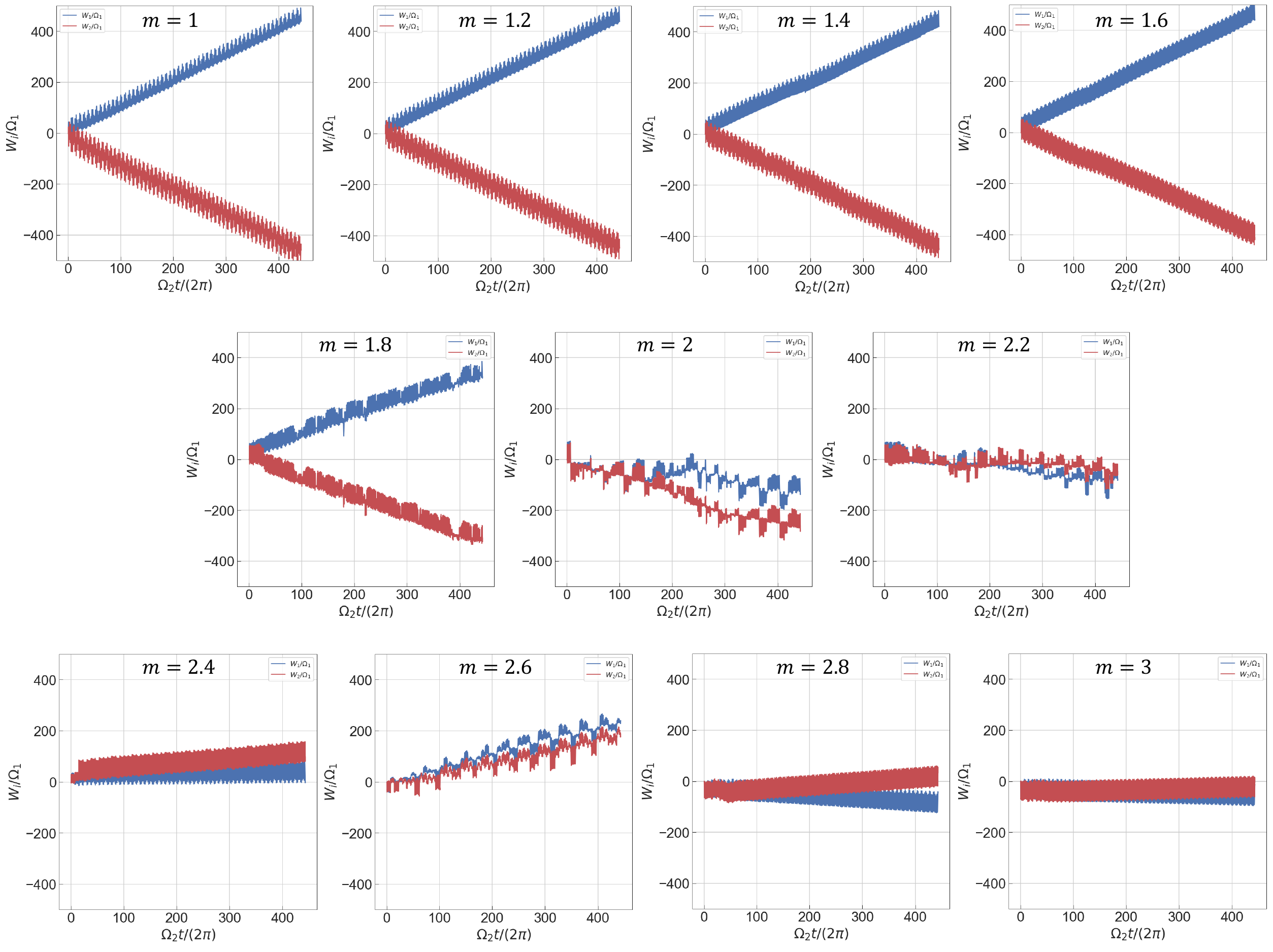}
        \caption{Topological energy pumping for $\Omega_2/\Omega_1 = \sqrt{\pi}$. The transition persists even for a different choice of irrational ratio. We simulate this with $gV_0 = 40\Omega_1$ and $\gamma = 0.01\Omega_1/\pi$.}
        \label{fig:rootpi_sweep}
    \end{figure}
    Floquet synthetic dimensions allow for the direct simulation of $k$-space Hamiltonians by emulating their evolution with incommensurate drive phases. Thus, we can achieve Hamiltonians such as
    \begin{equation}
        \mathcal{H}(\mathbf{k}) = \Omega_R\{\mathrm{sin}(k_x)\sigma_x+\mathrm{cos}(k_y)\sigma_y+\mathrm{cos}(k_z)\sigma_z\}
    \label{eq:Weyl_H}
    \end{equation}
    by simply performing the substitutions, $\Omega_1t + \phi_1 \rightarrow k_x$, $\Omega_2t + \phi_2 \rightarrow k_y$ and $\Omega_3t + \phi_3 \rightarrow k_z$. The linear evolution of these phases also emulates the effect of a synthetic electric field, which naturally allows us to study charge transport effects in these systems. We now study this Hamiltonian and its topological effects in detail. The gauge-invariant Berry curvature of this system can be written as
    \begin{equation}
        \mathbf{B}(\mathbf{k}) = \mathrm{Im}\left[ \frac{\bra{n(\mathbf{k})}\{\mathbf{\nabla}_\mathbf{k}\mathcal{H}(\mathbf{k})\}\ket{m(\mathbf{k})}\times\bra{m(\mathbf{k})}\{\mathbf{\nabla}_\mathbf{k}\mathcal{H}(\mathbf{k})\}\ket{n(\mathbf{k})}}{(E_m-E_n)^2} \right]
        \label{eq:Berry}
    \end{equation}
    We evaluate $\mathbf{\nabla}_\mathbf{k}\mathcal{H}(\mathbf{k})$ to be
    \begin{equation*}
        \mathbf{\nabla}_\mathbf{k}\mathcal{H}(\mathbf{k}) = \Omega_R[\hat{k}_x\cos(k_x)\sigma_x-\hat{k}_y\sin(k_y)\sigma_y-\hat{k}_z\sin(k_z)\sigma_z]
    \end{equation*}
    where $\hat{k}_i\,(i=x,y,z)$ is the unit vector in the Floquet momentum space. Another important quantity to calculate are the eigenvalues and eigenvectors of the Hamiltonian as functions of $\mathbf{k}$. Using the properties of spin-1/2 Hamiltonians, we get
    \begin{eqnarray}
        E_m &=& \Omega_R\sqrt{\sin^2k_x+\cos^2k_y+\cos^2k_z} = -E_n = E\nonumber\\
        \rho_m &=& |m\rangle\langle m| = \frac{1}{2}\left(\mathbb{I}+\frac{\mathcal{H}(\mathbf{k})}{E}\right)\nonumber\\
        \rho_n &=& |n\rangle\langle n| = \frac{1}{2}\left(\mathbb{I}-\frac{\mathcal{H}(\mathbf{k})}{E}\right)
        \label{eq:twoband}
    \end{eqnarray}
    Substituting these into Eq.~\ref{eq:Berry} and using properties of Pauli matrices,
    \begin{eqnarray}
        \mathbf{B}(\mathbf{k}) &= &\frac{\Omega_R^2}{4E^2}\mathrm{Im}\big[
        \hat{k}_x\sin(k_y)\sin(k_z)\bra{n}\sigma_y\rho_m\sigma_z-\sigma_z\rho_m\sigma_y\ket{n}-\hat{k}_y\cos(k_x)\sin(k_z)\bra{n}\sigma_z\rho_m\sigma_x-\sigma_x\rho_m\sigma_z\ket{n}\nonumber\\
        &&\;\;-\hat{k}_z\cos(k_x)\sin(k_y)\bra{n}\sigma_x\rho_m\sigma_y-\sigma_y\rho_m\sigma_x\ket{n}
        \big]\nonumber\\
        &=& \frac{\Omega_R^2}{8E^2}\mathrm{Im}\Bigg[
        \hat{k}_x\sin(k_y)\sin(k_z)\bra{n}\left([\sigma_y,\sigma_z]-2i\sin(k_x)\frac{\Omega_R\mathbb{I}}{E}\right)\ket{n}\nonumber\\
        &&\;\;-\hat{k}_y\cos(k_x)\sin(k_z)\bra{n}\left([\sigma_z,\sigma_x]-2i\cos(k_y)\frac{\Omega_R\mathbb{I}}{E}\right)\ket{n}\nonumber\\
        &&\;\;-\hat{k}_z\cos(k_x)\sin(k_y)\bra{n}\left([\sigma_x,\sigma_y]-2i\cos(k_z)\frac{\Omega_R\mathbb{I}}{E}\right)\ket{n}\Bigg]\nonumber\\
        &=& \frac{\Omega_R^2}{4E^2} \Bigg[
        \hat{k}_x\sin(k_y)\sin(k_z)\left(\bra{n}\sigma_x\ket{n}-\frac{\Omega_R\sin(k_x)}{E}\right)-\hat{k}_y\cos(k_x)\sin(k_z)\left(\bra{n}\sigma_y\ket{n}-\frac{\Omega_R\cos(k_y)}{E}\right)\nonumber\\
        &&\;\;-\hat{k}_z\cos(k_x)\sin(k_y)\left(\bra{n}\sigma_z\ket{n}-\frac{\Omega_R\cos(k_z)}{E}\right)\Bigg]
    \end{eqnarray}
    This is Eq.~\ref{eq:invariant} in the main text, and the Berry curvature in this form can be measured experimentally, when adiabaticity is maintained. Simplifying further, we get
    \begin{eqnarray}
        \mathbf{B}(\mathbf{k}) &= &\frac{\Omega_R^2}{4E^2} \Bigg[
        \hat{k}_x\sin(k_y)\sin(k_z)\left(\mathrm{Tr}\{\rho_n\sigma_x\}-\frac{\Omega_R\sin(k_x)}{E}\right)-\hat{k}_y\cos(k_x)\sin(k_z)\left(\mathrm{Tr}\{\rho_n\sigma_y\}-\frac{\Omega_R\cos(k_y)}{E}\right)\nonumber\\
        &&\;\;-\hat{k}_z\cos(k_x)\sin(k_y)\left(\mathrm{Tr}\{\rho_n\sigma_z\}-\frac{\Omega_R\cos(k_z)}{E}\right)\Bigg]\nonumber\\
        &=& \frac{-\hat{k}_x\sin(k_x)\sin(k_y)\sin(k_z)\sigma_x+\hat{k}_y\cos(k_x)\cos(k_y)\sin(k_z)\sigma_y+\hat{k}_z\cos(k_x)\sin(k_y)\cos(k_z)\sigma_z}{2\left(\sin^2k_x+\cos^2k_y+\cos^2k_z\right)^{3/2}}
    \end{eqnarray}
    where the final closed form expression for the Berry curvature comes from substituting Eq.~\ref{eq:twoband}, containing the expression for $\rho_n$. We then plot this in the main text to see the monopole behavior of Weyl points. The curvature at the Weyl point diverges, and integrating around this singularity leads to the quantized Chern number that characterizes the topological effects of this system.

    \begin{figure}
        \centering
        \includegraphics[width = 17cm]{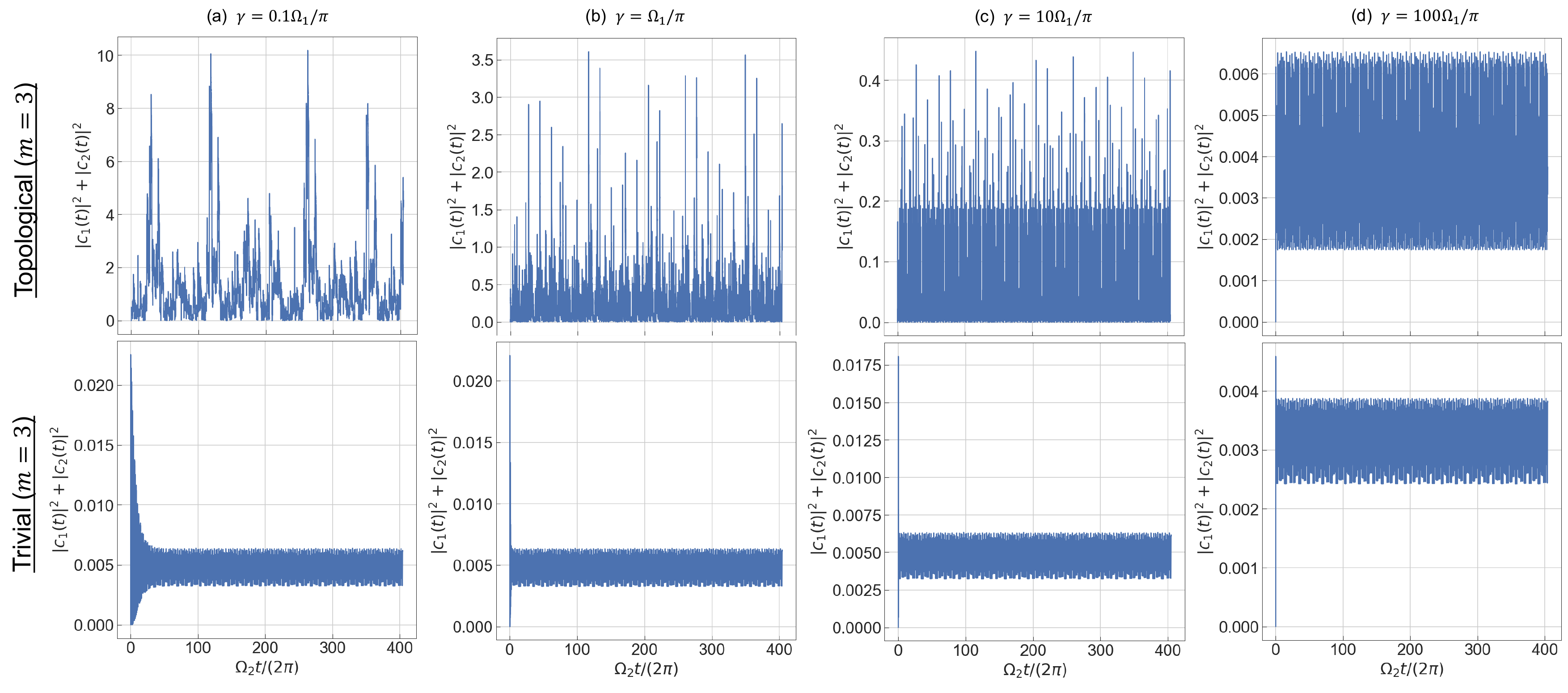}
        \caption{Total power in the photonic molecule for different photon loss rates. The driven-dissipative system reaches a quasi-steady state, with the intra-cavity power displaying oscillations, but not fully decaying or diverging.}
        \label{fig:total_power}
    \end{figure}

\end{document}